\documentclass[floatfix,superscriptaddress,nofootinbib,11pt]{revtex4-2}

\usepackage{naturehumbeh}

\newcommand{\samplesize}{6,890}

\begin{document}

%
%
%
%

\newcommand{\thetitle}{Political audience diversity and news reliability in algorithmic ranking}
\title{\thetitle{}}

\date{\today}

\newcommand{\usf}{Department of Computer Science and Engineering, University of South Florida, Tampa, FL, USA}
\newcommand{\dartmouth}{Department of Government, Dartmouth College, Hanover, NH, USA}
\newcommand{\sice}{School of Informatics, Computing, and Engineering, Indiana University Bloomington, Bloomington, IN, USA}
\newcommand{\iuni}{Network Science Institute, Indiana University, Bloomington, IN, USA}
\newcommand{\osome}{Observatory on Social Media, Indiana University, Bloomington, IN, USA}
\newcommand{\stanford}{Department of Political Science, Stanford University, Stanford, CA, USA}

\author{Saumya Bhadani}
\affiliation{\usf{}}

\author{Shun Yamaya}
\affiliation{\stanford{}}

\author{Alessandro Flammini}
\affiliation{\osome{}}

\author{Filippo Menczer}
\affiliation{\osome{}}

\author{Giovanni Luca Ciampaglia}
\email[Corresponding author. Email: ]{glc3@mail.usf.edu}
\affiliation{\usf{}}

\author{Brendan Nyhan}
\affiliation{\dartmouth{}}

\begin{abstract}
\noindent Newsfeed algorithms frequently amplify misinformation and other low-quality content. How can social media platforms more effectively promote reliable information? Existing approaches are difficult to scale and vulnerable to manipulation. In this paper, we propose using the political diversity of a website's audience as a quality signal. Using news source reliability ratings from domain experts and web browsing data from a diverse sample of \samplesize{} U.S. citizens, we first show that websites with more extreme and less politically diverse audiences have lower journalistic standards. We then incorporate audience diversity into a standard collaborative filtering framework and show that our improved algorithm increases the trustworthiness of websites suggested to users --- especially those who most frequently consume misinformation --- while keeping recommendations relevant. These findings suggest that partisan audience diversity is a valuable signal of higher journalistic standards that should be incorporated into algorithmic ranking decisions. 
\end{abstract}

\maketitle


\clearpage

\ifbool{misinforeview}{%
\section*{Implications (Why does this matter? And to whom?)}
}{%
}


\noindent Concerns continue to grow about the prevalence of misinformation on social media platforms~\citep{Lazer-fake-news-2018, vosoughi2018spread}, including during the recent COVID-19 pandemic~\cite{yang2020prevalence}. These types of content often exploit people's tendency to prefer pro-attitudinal information~\citep{hart2009feeling}, which can be exacerbated by platform content recommendations \citep{BakshyAdamic,chen2020neutral}. In this paper, we explore a possible algorithmic approach to mitigate the spread of misinformation and promote content with higher journalistic standards online. 

Social media platform recommendation algorithms frequently amplify bias in human consumption decisions. Though the information diets of Americans are less slanted in practice than many assume, the people who consume the most political news are most affected by the tendency toward selective exposure~\citep{guess18}. As a result, the news audience is far more polarized than the public as a whole~\citep{guessnd2, flaxman2016filter}. Although the prevalence of so-called ``fake news'' online is rather limited and concentrated among relatively narrow audiences~\cite{allcott2017social, guess18, grinberg2019fake, guess2019less, allen2020evaluating, guess2020exposure}, content that generally appeals to these tendencies --- which does include low-quality or false news --- may generate high levels of readership or engagement~\cite{vosoughi2018spread}, prompting algorithms that seek to maximize engagement to distribute them more widely.

Prior research indicates that existing recommendation algorithms tend to promote items that have already achieved popularity~\citep{goel2010anatomy,Nikolov2018biases}. This bias may have several effects on the consumption of low-quality and false news. First, sorting the news by engagement (either predicted or achieved) can exacerbate polarization by increasing in-group bias and discouraging consumption among outgroup members~\cite{shmargad2020sorting}. Second, it may contribute to information cascades, amplifying differences in rankings from small variations or random fluctuations and degrading the overall quality of information consumed by users~\citep{Salganik854, hogg2015disentangling,ciampaglia2018how, Germano:2019:FSC:3308558.3313693, Macyeaax0754}. Third, exposure to engagement metrics makes users more likely to share and less likely to fact-check highly engaging content from low-credibility sources, increasing vulnerability to misinformation~\cite{Fakey2020}. Finally, popularity bias in recommendation systems can create \emph{socio-algorithmic vulnerabilities} to threats such as automated amplifiers, which exploit algorithmic content rankings to spread low-quality and inflammatory content to like-minded audiences~\citep{shao2018spread,stella2018bots}. 

Given the speed and scale of social media, assessing directly the quality of every piece of content or the behavior of each user is infeasible. Online platforms are instead seeking to include signals about news quality in their content recommendation algorithms~\cite{fb2020prioritizing, google2020surfacing}, for example by extracting information from trusted publishers~\cite{shan2020factoring} or by means of linguistic patterns analysis~\cite{rashkin2017truth,shan2018linguistic}. More generally, a vast literature examines how to assess the credibility of online sources~\citep{gupta2014tweetcred,cho2015survey} and the reputations of individual online users~\citep{golbeck2005computing, Adler:2007:CRS:1242572.1242608}, which could in principle bypass the problem of checking each individual piece of content. Unfortunately, many of these methods are hard to scale to large groups and/or depend upon context-specific information about the type of content being generated. For example, methods for assessing the credibility of content on Wikipedia often assume content is organized as a wiki. As a result, they are not easily applied to news content recommendations on social media platforms. 

Another approach is to try to evaluate the quality of articles directly~\citep{zhang2018structured}, but scaling such an approach would likely be costly and cause lags in the evaluation of novel content. Similarly, while crowdsourced website evaluations have been shown to be generally reliable in distinguishing between high and low quality news sources~\citep{Pennycook2521}, the robustness of such signals to manipulation is yet to be demonstrated.

Building on the literature about the benefits of diversity at the group level~\cite{hong2004groups,shi2019wisdom}, we propose using the partisan diversity of the audience of a news source as a signal of its quality. This approach has two key advantages. First, audience partisan diversity can be computed at scale given that information about the partisanship of users is available or can be inferred in a reliable manner. Second, because diversity is a property of the audience and not of its level of engagement, it is less susceptible to manipulation if one can detect inauthentic partisan accounts~\citep{varol2017online,Yang2019botometer,Yang2020botometer-lite,botometerv4-2020}. These two conditions (inferring partisanship reliably and preventing abuse by automated amplification/deception) could easily be met by the major social media platforms, which have routine access to a wealth of signals about their users and their authenticity. 

We evaluate the merits of our proposed approach using data from two sources: a comprehensive data set of web traffic history from \samplesize{} 
Americans, collected along with surveys of self-reported partisan information from respondents in the YouGov Pulse survey panel, and a data set of 3,765 news source reliability scores compiled by trained experts in journalism and provided by NewsGuard~\cite{newsguard}. We first establish that domain pageviews are not associated with overall news reliability, highlighting the potential problem with algorithmic recommendation systems that rely on popularity and related metrics of engagement. We next define measures of audience partisan diversity and show that these measures correlate with news reliability better than popularity does. Finally, we study the effect of incorporating audience partisan diversity into algorithmic ranking decisions. When we create a variant of the standard collaborative filtering algorithm that explicitly takes audience partisan diversity into account, our new algorithm provides more trustworthy recommendations than the standard approach with only a small loss of relevance, suggesting that reliable sources can be recommended without the risk of jeopardizing user experience. 

These results demonstrate that diversity in audience partisanship can serve as a useful signal of news reliability at the domain level, a finding that has important implications for the design of content recommendation algorithms used by online platforms. Although the news recommendation technologies deployed by platforms are more sophisticated than the approach tested here, our results highlight a fundamental weakness of algorithmic ranking methods that prioritize content that generates engagement and suggest a new metric that could help improve the reliability of the recommendations that are provided to users. 

\ifbool{misinforeview}{\section*{Findings}}{}%
\ifbool{naturehumbeh}{\section*{Results}}{}%
\label{sec:results}

\subsection*{Popularity does not predict news reliability}

\begin{figure}
\includegraphics[width=\textwidth]{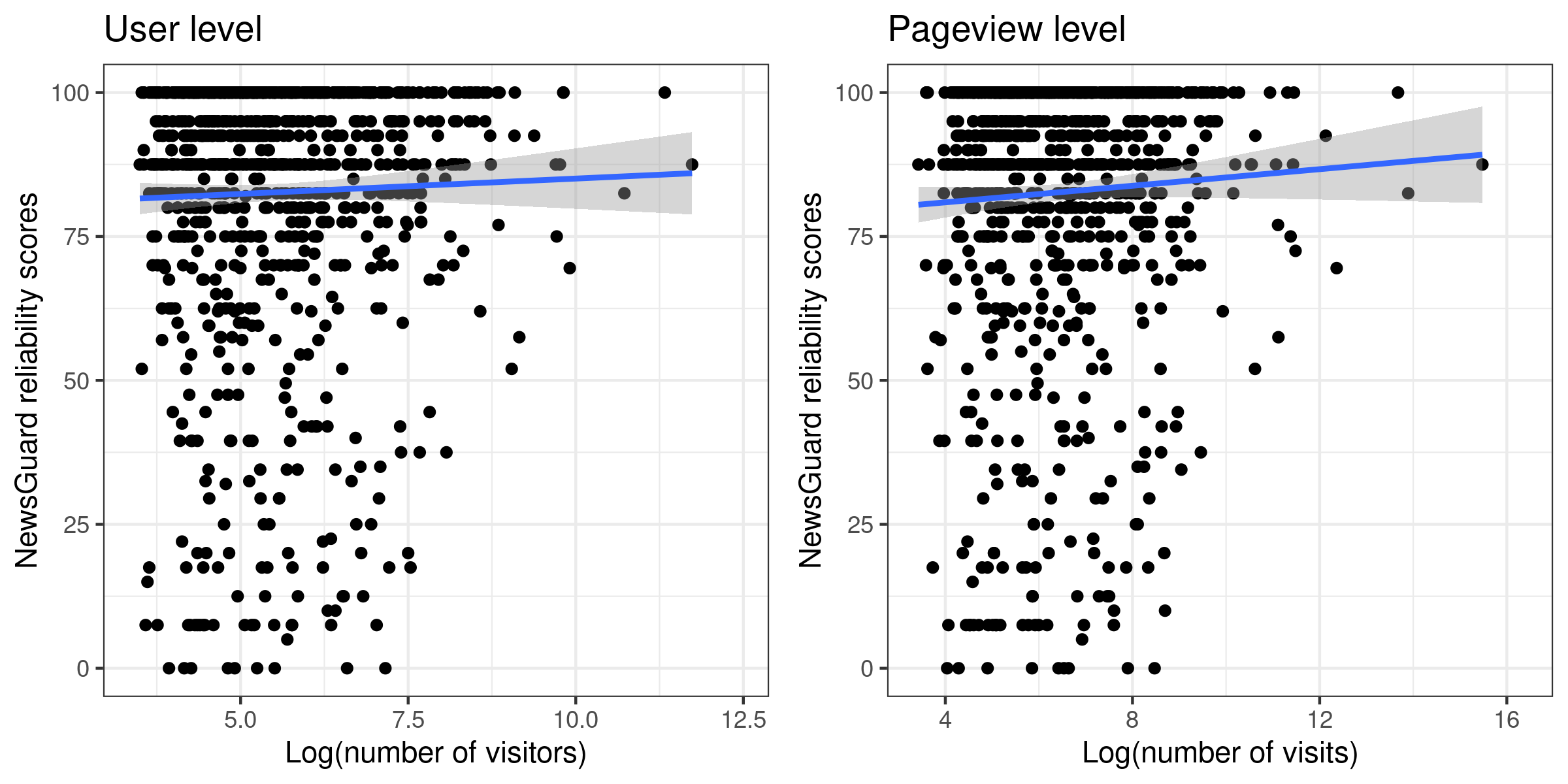}
\caption{Relationship between audience size and news reliability by domain. Reliability scores provided by NewsGuard~\cite{newsguard}.}
\label{fig:traffic_vs_quality}
\end{figure}

To motivate our study, we first demonstrate that the popular news content that algorithmic recommendations often highlight is not necessarily reliable. To do so, we assess the relationship between source popularity and news reliability. We measure source popularity using the YouGov Pulse traffic data. Due to skew in audience size among domains, we transform these data to a logarithmic scale. In practice, we measure the popularity of a source in two ways: as the (log of) number of users, and as the (logged) number of visits, or pageviews. News reliability is instead measured using NewsGuard scores (see Methods~\ref{sec:data}). 
Fig.~\ref{fig:traffic_vs_quality} shows that the popularity of a news source is at best weakly associated with its reliability. At the user level (left pane),
the overall Pearson correlation is $r=0.03$ (two-sided $p = 0.36$). At the pageview level (right pane), $r = 0.05$ (two-sided $p = 0.12$). The association between the two variables remains weak even if we divide sources based on their partisanship.
When measuring popularity at the user level, websites that have a predominantly Democratic audience have a significant positive association ($r=0.09$, two-sided $p = 0.02$), but for websites with a Republican audience the correlation is negative and not significant at conventional standards ($r = -0.12$, two-sided $p = 0.06$). A similar pattern holds at the pageview level: a weak positive association for websites with predominantly Democratic audiences ($r=0.08$, two-sided $p=0.02$) and a negative but not significant association for those with predominantly Republican audiences ($r=-0.06$, two-sided $p=0.34$). Overall, these results suggest the strength of association between the two variables is quite weak.

%

\subsection*{Audience partisan diversity is a signal for high-reliability news}

\begin{figure}
\includegraphics[width=\textwidth]{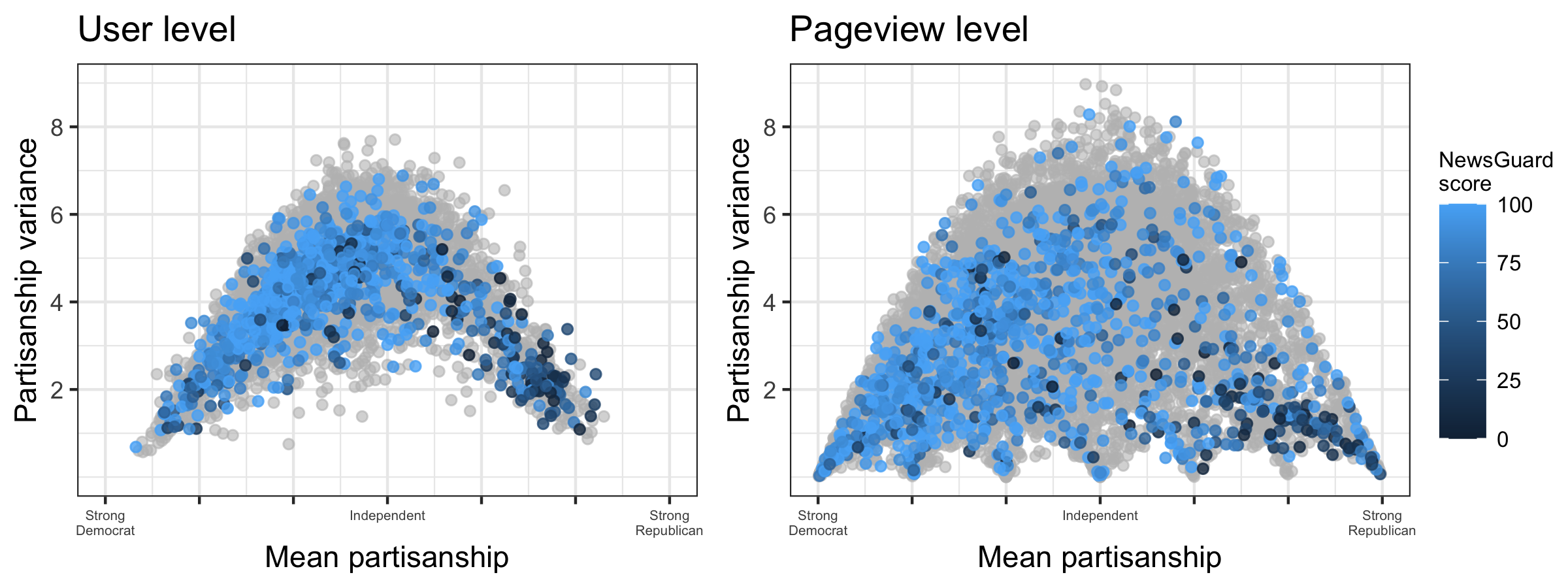}
\caption{Average audience partisanship versus variance.
Left panel: user level. Right panel: pageview level. Domains for which we have NewsGuard reliability scores~\cite{newsguard} are shaded in blue (where darker shades equal lower scores). Domains with no available score are plotted in gray.}
\label{fig:average_slant_vs_variance1}
\end{figure}

In contrast, we observe that sites with greater audience partisan diversity tend to have higher NewsGuard scores while those with lower levels of diversity, and correspondingly more homogeneous partisan audiences, tend to have lower reliability scores. As our primary metric of diversity, we selected from a range of alternative definitions (see Methods~\ref{sec:ad}) the variance of the partisanship distribution. Fig.~\ref{fig:average_slant_vs_variance1} shows how NewsGuard scores vary with both mean audience partisanship and the variance in audience partisanship.

As Fig.~\ref{fig:average_slant_vs_variance1} indicates, unreliable websites with very low NewsGuard scores are concentrated in the tails of the distribution, where partisanship is most extreme and audience partisan diversity is, by necessity, very low. This relationship is not symmetrical: low-reliability websites (whose markers are darker shades of blue in the figure) are especially concentrated in the right tail, which corresponds to websites with largely Republican audiences. The data in Fig.~\ref{fig:average_slant_vs_variance1} also suggests that the reliability of a website may be associated not just with the variance of the distribution of audience partisanship slants, but also with its mean. To account for this, we first compute the coefficient of partial correlation between NewsGuard reliability scores and the variance of audience partisanship given the mean audience partisanship of each website. Compared with popularity, we find a stronger (and significant) correlation regardless of whether mean partisanship and audience partisan diversity are calculated by weighting individual audience members equally (user level, left panel: partial correlation $r = 0.38$, two-sided $p < 10^{-4}$) or by how often they visited a given site (pageview level, right panel: partial correlation $r = 0.22$, two-sided $p < 10^{-4}$). 
 
Aside from mean partisanship, a related, but potentially distinct, confounding factor is the extremity of the partisanship slants distribution (i.e., the distance of the average partisanship of a website visitor on a 1--7 scale from the midpoint of 4, which represents a true independent).  We thus computed partial correlation coefficients again, but instead keep the ideological extremity of website audiences constant instead of the mean. Our results are consistent using this approach (user level: $r = 0.26, p < 10^{-4}$; pageview level: $r=0.15, p < 10^{-4}$; both tests are two-sided).

\begin{figure}
\includegraphics[width=\textwidth]{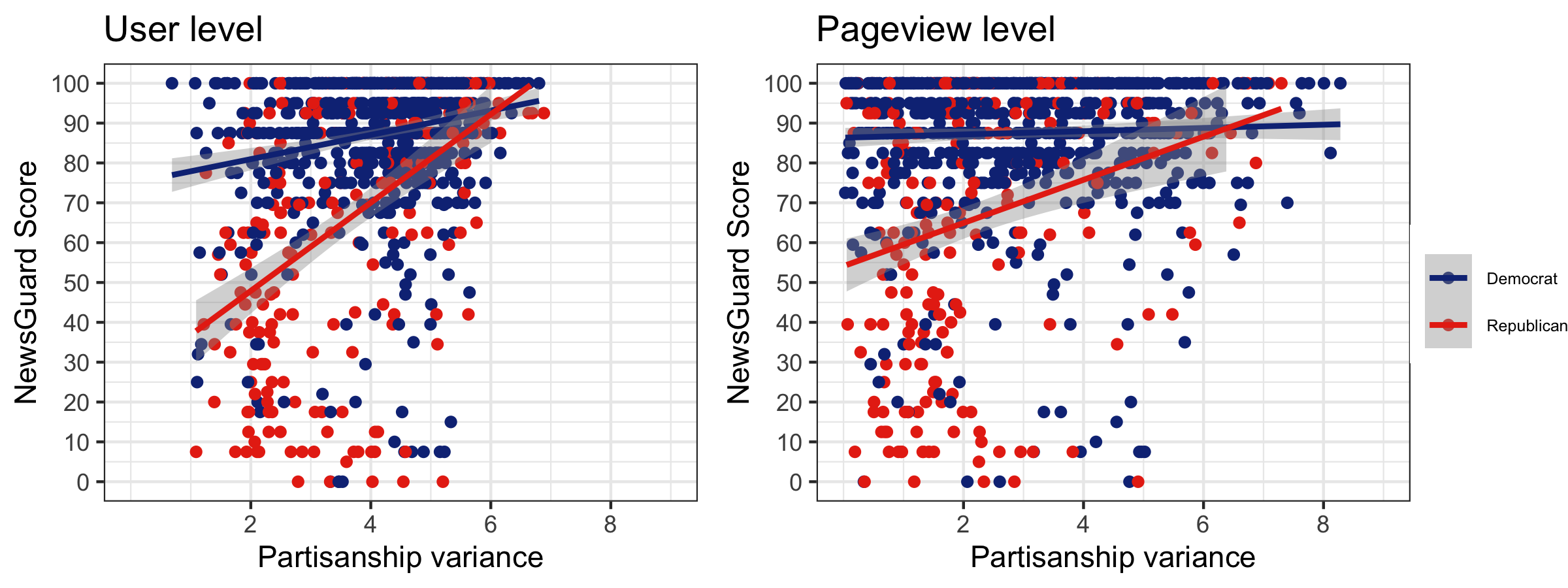}
\caption{Relationship between audience partisan diversity and news reliability for websites whose average visitor is a Democrat or a Republican. Left panel: variance computed at user level. Right panel: variance computed at pageview level. News reliability scores from NewsGuard~\cite{newsguard}.}
\label{fig:average_slant_vs_variance2}
\end{figure}

We study the diversity--reliability relationship in more detail in Fig.~\ref{fig:average_slant_vs_variance2}, which differentiates between websites with audiences that are mostly Republican and those with audiences that are mostly Democratic. Consistent with what we report above, Fig.~\ref{fig:average_slant_vs_variance2} shows that audience partisan diversity is positively associated with news reliability. Again, this relationship holds both when individual audience members are weighted equally (user level, left panel) and when they are weighted by their number of accesses (pageview level, right panel), though the association is stronger at the user level (standardized OLS coefficient: $\beta = 6.67\,(0.58)$ at user level; $\beta = 3.91\,(0.71)$ at pageview level). In addition, we find that the relationship is stronger for sites whose average visitor identifies as a Republican (standardized OLS coefficient of Republican domains: $\beta = 13.1\,(1.59)$ at user level; $\beta = 9.61\,(2.20)$ at pageview level) versus those whose average visitor identifies as a Democrat (standardized OLS coefficient of Democrat domains: $\beta = 2.77\,(0.54)$ at user level; $\beta = 0.68\,(0.61)$ at pageview level), which is consistent with Fig.~\ref{fig:average_slant_vs_variance1}. 

These results are not affected by popularity. Partisan diversity is weakly correlated with popularity, regardless of the operational definition of either measure (see Supplementary Materials). In fact, the association between diversity and Newsguard reliability scores is consistent even when controlling for popularity (user level: $r = 0.34$, two-sided $p < 10^{-4}$; pageview level: $r = 0.17$, two-sided $p < 10^{-4}$), suggesting that diversity could contribute to detecting quality over and above the more typical popularity metrics used by social media algorithms. However, the previous analysis of Fig.~\ref{fig:average_slant_vs_variance2} shows that the overall relationship masks significant heterogeneity between websites with mostly Republican or Democratic audiences. To tease apart the contributions of popularity from those of partisanship, we estimate a full multivariate regression model. After controlling for both popularity and political orientation, we find qualitatively similar results. Full regression tables can be found in Supplementary Materials. 

As mentioned before, variance in audience partisanship is not the only possible way to define audience partisan diversity; alternative definitions can be used (e.g., entropy; see Methods~\ref{sec:ad}). As a robustness check, we therefore consider a range of alternative definitions of audience partisan diversity and obtain results that are qualitatively similar to the ones presented here, though results are strongest for variance (see Supplementary Materials). 

\subsection*{Audience partisan diversity produces trustworthy, relevant recommendations}

To understand the potential effects of incorporating audience partisan diversity into algorithmic recommendations, we next consider how recommendations from a standard user-based collaborative filtering (CF) algorithm~\cite{resnick1994grouplens, konstan1997grouplens} change if we include audience partisan diversity as an additional signal. We call this modified version of the algorithm CF+D, which stands for Collaborative Filtering~+~Diversity (see Methods~\ref{sec:cf+d} for formal definition). 

In classic CF, users are presented with recommendations drawn from a set of items (in this case, web domains) that have been ``rated'' highly by those other users whose tastes are most similar to theirs. Lacking explicit data about how a user would ``rate'' a given web domain, we use a quantity derived from the number of user pageviews to a domain (based on TF-IDF; see also Methods~\ref{sec:cf+d}) as the rating.

To evaluate our method, we follow a standard supervised learning workflow. We first divide web traffic data for each user in the YouGov Pulse panel into training and testing sets by domain (see Methods~\ref{sec:supervised}). We then compute similarities in traffic patterns between users for all domains in the training set (not just news websites) and use the computed similarities to predict the aforementioned domain-level pageviews metric on the test set. The domains that receive the highest predicted ratings (i.e., expected TF-IDF-transformed pageviews) are then selected as recommendations. As a robustness check, we obtain consistent results if we split the data longitudinally instead of randomly (i.e., as a forecasting exercise; see Supplementary Materials for details).  

Note that if a user has not visited a domain, then the number of visits for that domain will be zero. In general, due to the long tail in user interests~\cite{goel2010anatomy}, we cannot infer that the user has a negative preference toward a website just because they have not visited it. The user may simply be unaware of the site. We therefore follow standard practice in the machine learning literature in only evaluating recommendations for content for which we have ratings (i.e., visits in the test set), though in practice actual newsfeed algorithms rank items from a broader set of inputs, which typically includes content the user may not have seen (for example, content shared by friends~\cite{BakshyAdamic}).

To produce recommendations for a given user, we consider all the domains visited by the user in the test set for which ratings are available from one or more respondents in a neighborhood of most similar users (domains with no neighborhood rating are discarded since neither CF nor CF+D can make a prediction for them; see Methods~\ref{sec:cf+d}) and for which we have a NewsGuard score (i.e., a reliability score). We then rank those domains by their rating computed using either CF or CF+D. 
This process produces a ranked list of news domains and reliability scores from both the standard CF algorithm and the CF+D algorithm, which has been modified to incorporate the audience partisan diversity signal. We evaluate these lists using two different measures of trustworthiness which are computed for the top $k$ domains in each list: the mean score (a number in the 0--100 range) and the proportion of domains with a score of 60 or higher, which NewsGuard classifies as indicating that a site ``generally adheres to basic standards of credibility and transparency''~\cite{newsguard} (see Methods~\ref{sec:trust}).

\begin{figure}
      \includegraphics[width=0.495\textwidth]{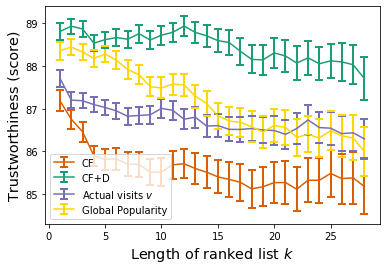}
      \includegraphics[width=0.495\textwidth]{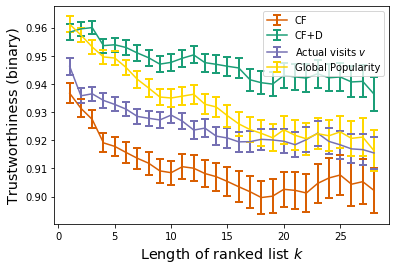}
    \caption{Trustworthiness of recommended domains by length of ranked list $k$. Left: Trustworthiness based on scores from NewsGuard~\cite{newsguard}. Right: proportion of domains labeled as `trustworthy', also by NewsGuard. Actual visits $v$ are normalized using TF-IDF (see Methods~\ref{sec:cf+d}). Global popularity is overall domain popularity (see Methods~\ref{sec:global_popularity}). Each bin represents the average computed on the top-$k$ recommendations for all users in the YouGov panel with $\ge k$ recommendations in their test sets. Bars represent the standard error of the mean. The values of $k$ are capped so that each bin has $\ge 100$ users in it (see Supplementary Materials for plot with all values of $k$). In this figure, both CF and CF+D compute the similarity between users using the Kendall $\tau$ correlation coefficient (see Methods~\ref{sec:cf+d}). We obtain qualitatively similar results using the Pearson correlation coefficient (see Supplementary Materials).}
    \label{fig:trust}
\end{figure}

By varying the number of top domains $k$, we can evaluate how trustworthiness changes as the length of the list of recommendations increases. In Fig.~\ref{fig:trust} we plot the trustworthiness of the recommended domains as a function of $k$. We restrict values of $k$ to 1--28, the values for which there are at least $100$ users in each bin (plots spanning the full range are available in Supplementary Materials). Each panel compares the average trustworthiness of domains ranked by CF and CF+D with two baselines. The first is the trustworthiness of websites users visited in the test set ranked by their TF-IDF-transformed number of visits (i.e.,~pageviews). This baseline captures the trustworthiness of the websites that users in the YouGov Pulse panel actually visited after adjusting for the fact that more popular websites tend to attract more visits in general. The second baseline is the trustworthiness of recommendations produced according to the overall popularity of domains. This baseline does not include any local information about user-user similarities, and thus can be seen as a ``global'' measure of popularity with no contribution due to user personalization (see Methods~\ref{sec:global_popularity}).

We observe in Fig.~\ref{fig:trust} that the trustworthiness of recommendations produced by CF+D is significantly better than standard CF recommendations, global popularity recommendations, and baseline statistics from user behavior. In particular, CF produces less trustworthy rankings than both the recommendations based on global popularity and on user visits (for small values of $k$ the difference is within the margin of error). In contrast, CF+D produces rankings that are more trustworthy than CF and either baseline (global popularity or actual visits) across different levels of $k$. These results suggest that audience partisan diversity can provide a valuable signal to improve the reliability of algorithmic recommendations.

Of course, the above exercise would be meaningless if our proposed algorithm recommended websites that do not interest users. Because CF+D alters the set of recommended domains to prioritize those visited by more diverse partisan audiences, it may be suggesting sources that offer counter-attitudinal information or that users do not find relevant. In this sense, CF+D could represent an audience-based analogue of the topic diversification strategy from the recommender systems literature~\cite{ziegler2005improving}. If so, a loss of predictive ability would be expected.

\begin{figure}
    \includegraphics[width=0.495\textwidth]{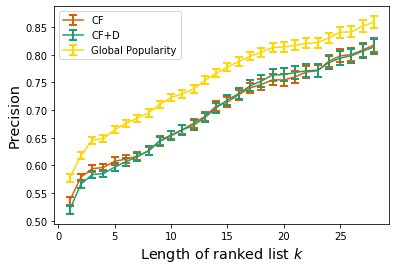}
    \includegraphics[width=0.495\textwidth]{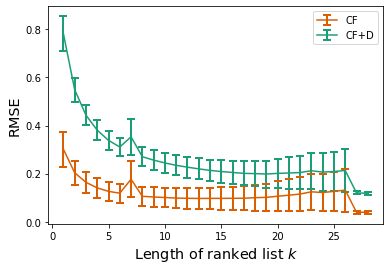}
    \caption{Accuracy of domain recommendations by length of ranked list $k$. Left: Precision (proportion of correctly ranked sites) by length of ranked list $k$ (higher is better). Right: RMSE (root mean squared error) of predicted pageviews for top $k$ ranked domains by length of ranked list $k$ (lower is better). Each bin represents the average computed on the top-$k$ recommendations of all users with $\ge k$ recommendations in their test sets. Bars represent the standard error of the mean. The values of $k$ are capped so that each bin has $\ge 100$ users in it (see Supplementary Materials for plot with all values of $k$). In this figure, both CF and CF+D compute the similarity between users using the Kendall $\tau$ correlation coefficient (see Methods~\ref{sec:cf+d}). We obtain qualitatively similar results using the Pearson correlation coefficient (see Supplementary Materials).}
    \label{fig:precision}
\end{figure}

Fig.~\ref{fig:precision} compares the accuracy of CF+D in predicting user visits to domain in the test set with that of CF. To evaluate accuracy, we compute both the fraction of correctly predicted domains (precision) and root mean squared error (RMSE) as a function of the number of recommended domains $k$ (see Methods~\ref{sec:eval} for definitions). Note that precision improves with $k$ (left panel) by definition --- as $k$ grows, we are comparing an increasingly large set of recommendations with a list of fixed size. Because each bin averages over users with at least $k$ domains in their test set, when $k$ reaches the maximum size of the recommendation list we can make, the precision necessarily becomes 100\%. Note that the plots in Fig.~\ref{fig:precision} do not reach this level --- they include only bins with at least $100$ users in them --- but trend upward with $k$. (In the Supplementary Materials, we show plots that include results for all values of $k$.) 

As with precision, RMSE declines with $k$ (right panel) since we focus progressively on users with longer lists and thus more training data. Like in the left panel, each bin in the right panel averages over users with at least $k$ domains in their test set. Unlike precision, however, RMSE is more prone to producing outliers because it does not depend on the relative ranking of item ratings but instead on their magnitude. This difference is reflected in the sudden drop in the error bars for the RMSE at $k=27$ due to the presence of a single user with a maximum list length of 26 domains in testing. We manually checked the data of this user and found that the training set included only domains visited infrequently, leading to large errors. Removing this outlier eliminated the observed change. 


To provide intuition about the contribution of popularity in recommendations, the left panel of Fig.~\ref{fig:precision} also shows the precision of the na\"{\i}ve baseline obtained by ranking items by their global popularity. This baseline outperform CF and CF+D but at the price of providing the same set of recommendations to all users (i.e., the results are not personalized) and of providing recommendations of lower trustworthiness (Fig.~\ref{fig:trust}). Note that the RMSE cannot be computed for this baseline because this metric requires knowledge of the rating of a domain, not just of its relative ranking.

Our results are generally encouraging. In both cases, precision is low and RMSE is high for low values of $k$, but error levels start to stabilize around $k = 10$, which suggests that making correct recommendations for shorter lists (i.e., $k< 10$) is more challenging than for longer ones. Moreover, when we compare CF+D with CF, accuracy declines slightly for CF+D relative to CF but the difference is not statistically significant for all but small values of $k$, suggesting that CF+D is still capable of producing relevant recommendations. 

Re-ranking items by diversity has minimal effects on predictive accuracy, but how does it affect user satisfaction? The recommendations produced by CF+D would be useless if users did not find them engaging. Unfortunately, we lack data about user satisfaction in the YouGov panel - our primary metric (log number of website visits) cannot be interpreted as a pure measure of satisfaction (other factors of course shape the decision by users in the YouGov panel to visit a website, including social media recommendations themselves). 

However, it is possible that more accurate recommendations will result in higher user satisfaction. In the revised version, we therefore quantify the significance of the observed drop in accuracy due to re-ranking by diversity. More specifically, we simulated the sampling distribution of the precision of recommendation after re-ranking by re-shuffling domain labels in the list of ratings produced by CF+D. This procedure allows us to calculate the probability of a drop in precision as small as the observed one due to random chance alone. Compared with this null model, we find that our results lead to significantly higher precision --- most random re-rankings of the same magnitude as the one produced by CF+D would result in lower precision than what we observe. We report the results of this additional analysis in the Supplementary Materials (Fig.~\ref{fig:resampling}).

\subsection*{Audience partisan diversity mitigates selective exposure to misinformation}

The results above demonstrate that incorporating audience partisan diversity can increase the trustworthiness of recommended domains while still providing users with relevant recommendations. However, we know that exposure to unreliable news outlets varies dramatically across the population. For instance, exposure to untrustworthy content is highly concentrated among a narrow subset of highly active news consumers with heavily slanted information diets~\cite{grinberg2019fake,guess2020exposure}. We therefore take advantage of the survey and behavioral data available on participants in the Pulse panel to consider how CF+D effects vary by individual partisanship (self-reported via survey), behavioral measures such as volume of news consumption activity and information diet slant, and contextual factors that are relevant to algorithm performance such as similarity with other users.

In this section, we again produce recommendations using either CF or CF+D and measure their difference in trustworthiness with respect to a baseline based on user visits (specifically the ranking by TF-IDF-normalized number of visits $v$; see Methods~\ref{sec:cf+d}). However, we analyze the results differently than those reported above. Rather than considering recommendations for lists of varying length $k$, we create recommendations for different subgroups based on the factors of interest and compare how the effects of the CF+D approach vary between those groups. 

To facilitate comparisons in performance between subgroups that do not depend on list length $k$, we define a new metric to summarize the overall trustworthiness of the ranked lists obtained with CF and CF+D over all possible values of $k$. Since users tend to pay less attention to items ranked lower in the list~\cite{10.1145/3130332.3130334}, it is reasonable to assume that lower-ranked items ought to contribute less to the overall trustworthiness of a given ranking. 

Let us now consider probabilistic selections from two different rankings, represented by random variables $X$ and $X'$, where $X$ is the random variable of the ranking produced by one of the two recommendation algorithms (either CF or CF+D) and $X'$ is the selection from the baseline ranking based on user visits. Using a probabilistic discounting method (see Eq.~\ref{eq:ranking} in Method~\ref{sec:rankingmodel}), we compute the expected change in trustworthiness $Q$ from switching the selection from $X'$ to $X$,
\begin{linenomath*}
\begin{equation}\label{eq:change}
\Delta Q = \mathsf{E}\left[Q(X)\right] - \mathsf{E}\left[Q(X')\right]
\end{equation}
\end{linenomath*}
where the expectations of $Q(X)$ and $Q(X')$ are taken with regard to the respective rankings (see Methods~\ref{sec:rankingmodel}). A value of $\Delta Q > 0$ indicates that algorithmic recommendations are more trustworthy than what users actually accessed. If $\Delta Q < 0$, the trustworthiness of a ranked list is lower than the baseline from user visits. (To ensure that the results below are not affected by the discounting method we employ, we report qualitatively similar results obtained without any discounting for a selection of values of $k$ in the Supplementary Materials.)

\begin{figure} 
\includegraphics[width=\textwidth]{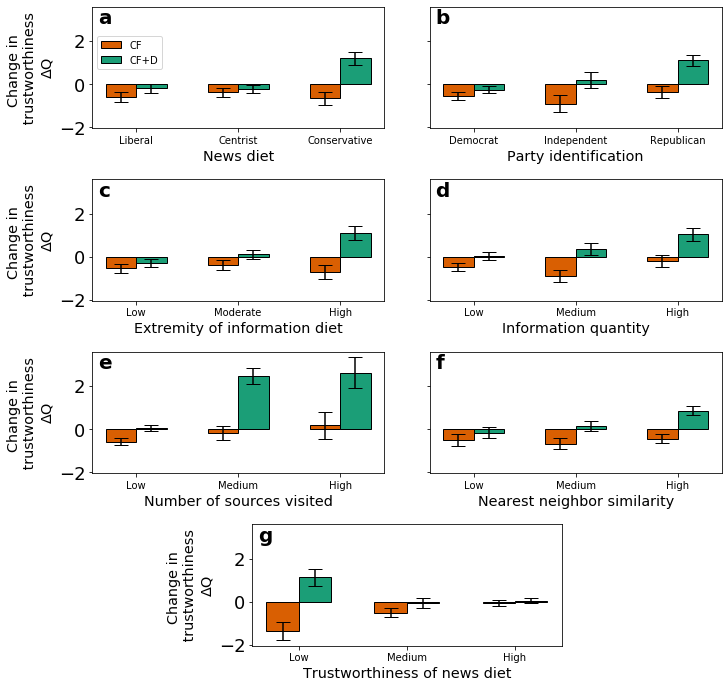}
\caption{Effect of CF and CF+D (versus actual visits baseline) on trustworthiness by user characteristics and behavior. (a)~Ideological slant of visited domains (terciles using scores from \citeauthor{BakshyAdamic}~\cite{BakshyAdamic}). (b)~Self-reported party ID from YouGov Pulse responses as measured on a 7-point scale (1--3: Democrats including people who lean Democrat but do not identify as Democrats, 4: Independents, 5--7: Republicans including people who lean Republican but do not identify as Republicans). (c)~Absolute slant of visited domains (terciles using scores from \citeauthor{BakshyAdamic}). (d)~Total online activity (TF-IDF-transformed pageviews; terciles). (e)~Distinct number of domains visited (terciles). (f)~Average user-user similarity with nearest $n=10$ neighbors in training set (terciles) (g)~Trustworthiness of domains visited by users (in training set; terciles). Bars represent the standard error of the mean of each stratum. Change in trustworthiness $\Delta Q$ based on scores from NewsGuard~\cite{newsguard}.}
\label{fig:tw_vs}
\end{figure}

Applying Eq.~\ref{eq:change}, we find that CF+D substantially increases trustworthiness
for users who tend to visit sources that lean conservative (Fig.~\ref{fig:tw_vs}(a)) and for those who have the most polarized information diets (in either direction; see Fig.~\ref{fig:tw_vs}(c)), two segments of users who are especially likely to be exposed to unreliable information~\cite{allcott2017social,grinberg2019fake,guess2020exposure}. In both cases, CF+D achieves the greatest improvement among the groups where CF reduces the trustworthiness of recommendations the most, 
which highlights the pitfalls of algorithmic recommendations for vulnerable audiences and the benefits of prioritizing sources with diverse audiences in making recommendations to those users.

Note that even though the YouGov sample includes self-reported information on both party ID and partisanship of respondents. We use only the former (Fig.~\ref{fig:tw_vs}(b)) for stratification to avoid circularity given the definition of CF+D, which relies on the latter. In Figs~\ref{fig:tw_vs}(a) and~\ref{fig:tw_vs}(c), we instead stratify on an external measure of news diet slant (calculated from a large sample of social media users; see Methods~\ref{sec:stratification}).

We also observe that CF+D has strong positive effects 
for users who identify as Republicans or lean Republican (Fig.~\ref{fig:tw_vs}(b)) and for those who are the most active news consumers in terms of both total consumption (Fig.~\ref{fig:tw_vs}(d)) and number of distinct sources (Fig.~\ref{fig:tw_vs}(e)). Furthermore, since the two recommendation schemes considered here (CF and CF+D) are predicated on identifying similar users according to their tastes and behaviors, we also segment the users of the YouGov sample according to the degree of similarity with their nearest neighbors (identified based on Kendall's rank correlation coefficient between user vectors; see Methods~\ref{sec:cf+d}). Stratifying on the average of nearest neighbor similarities, we find that CF+D results in improvements for the users whose browsing behavior is most similar to others in their neighborhood and who might thus be most at risk of ``echo chamber'' effects (Fig.~\ref{fig:tw_vs}(f)). Finally, when we group users by the trustworthiness of the domains they visit, we find that the greatest improvements from the CF+D algorithm occur for users who are exposed to the least trustworthy information (Fig.~\ref{fig:tw_vs}(g)). By contrast, the standard CF  algorithm often recommends websites that are less trustworthy than those that respondents actually visit ($\Delta Q < 0$). 

\section*{Discussion}

The findings presented here suggest that the ideological diversity of the audience of a news source is a reliable indicator of its journalistic quality. To obtain these findings, we combined source reliability ratings compiled by expert journalists with traffic data from the YouGov Pulse panel. Of course, we are not the first to study the information diets of Internet users. Prior work has leveraged Web traffic data to pursue related topics such as identifying potential dimensions of bias of news sources~\cite{ribeiro2018media, Nikolov2018biases}, designing methods to present diverse political opinions~\cite{munson2010presenting, munson2013encouraging}, and measuring the prevalence of filter bubbles~\cite{flaxman2016filter}. Unlike these studies, however, we focus on how to promote exposure to trustworthy information rather than seeking to quantify or reduce different sources of bias. 

A number of limitations must be acknowledged. First, our current methodology, which is based on reliability ratings compiled at the level of individual sources, does not allow us to evaluate the quality of specific articles that participants saw. However, even a coarse signal about source quality could still be useful for ranking a newsfeed given that information about reliability is more widely available at the publisher level than the article level. Another limitation is that our data lack information about actual engagement. Though we show  that our re-ranking procedure is associated with a minimal loss in predictive accuracy, it remains an open question whether diversity-based rankings lead not just to higher exposure with trustworthy content, but also to more engagement with it. More research is needed to tease apart the causal link between political attitudes, readership, engagement, and information quality.

Our work has a number of implications for the integrity of the online information ecosystem. First, our findings suggest that search engines and social media platforms should consider including audience diversity to their existing set of news quality signals. 
%
%
Such a change could be especially valuable for domains for which we lack other signals about their quality like source reliability ratings compiled by experts. Media ratings systems such as NewsGuard could also benefit from adopting our diversity metric, for example to help screen and prioritize domains for manual evaluation.

\begin{figure}
    \includegraphics[width=0.5\textwidth]{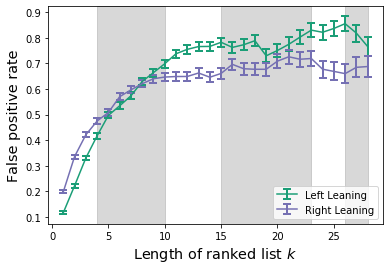}
    \caption{Probability that a trustworthy domain (NewsGuard score $\ge 60$) is not recommended by CF+D but is recommended by CF for left- and right-leaning domains. Each point is the average over a sample of users. The shaded regions represent the values of $k$ for which the difference is not statistically significant at standard levels ($\alpha = 0.05$, Welch's $t$-tests with Bonferroni correction for $n = 28$; all tests are two-sided).}
    \label{fig:false_pos_fig}
\end{figure}

Critics may raise concerns that such a change in ranking criteria would result in unfair outcomes, for example by reducing exposure to content by certain partisan groups but not others. To see whether ranking by diversity leads to any differential treatment for different partisan news sources, we compute the rate of \emph{false positives} due to re-ranking by diversity. Here the false positive rate is defined as the conditional probability that CF+D does not rank a trustworthy domain among the top $k$ recommendations while CF does for both left- and right-leaning domains. To determine whether a domain is trustworthy we rely on the classification provided by NewsGuard (i.e. the domain has a reliability score $\ge 60$). Fig.~\ref{fig:false_pos_fig} shows the rate of false positives as a function of $k$ of both left- and right-leaning domains averaged over all users. Despite some small differences, especially for low values of $k$, we find no consistent evidence that this change would produce systematically differential treatment across partisan groups. 

Another concern is the possibility of abuse. For example, an attacker could employ a number of automated accounts to collectively engage with an ideologically diverse set of sources. This inauthentic, ideologically diverse audience could then be used to push specific content the attacker wants to promote atop the rankings of a recommender system. Similarly, an attacker who wanted to demote a particular content could craft an inauthentic audience with low diversity. Fortunately, there is a vast literature on the topic of how to defend recommender systems against such ``shilling'' attacks~\cite{lam2004shilling,gunes2014shilling} and platforms already collect a wealth of signals to detect and remove inauthentic coordinated behavior of this kind. Future work should investigate the feasibility of creating trusted social media audiences that are modeled on existing efforts in marketing research using panels of consumers. We hope that our result stimulates further research in this area.

\section*{Methods}

\subsection{Data} 
\label{sec:data}

\begin{figure}
\includegraphics[width=.75\textwidth]{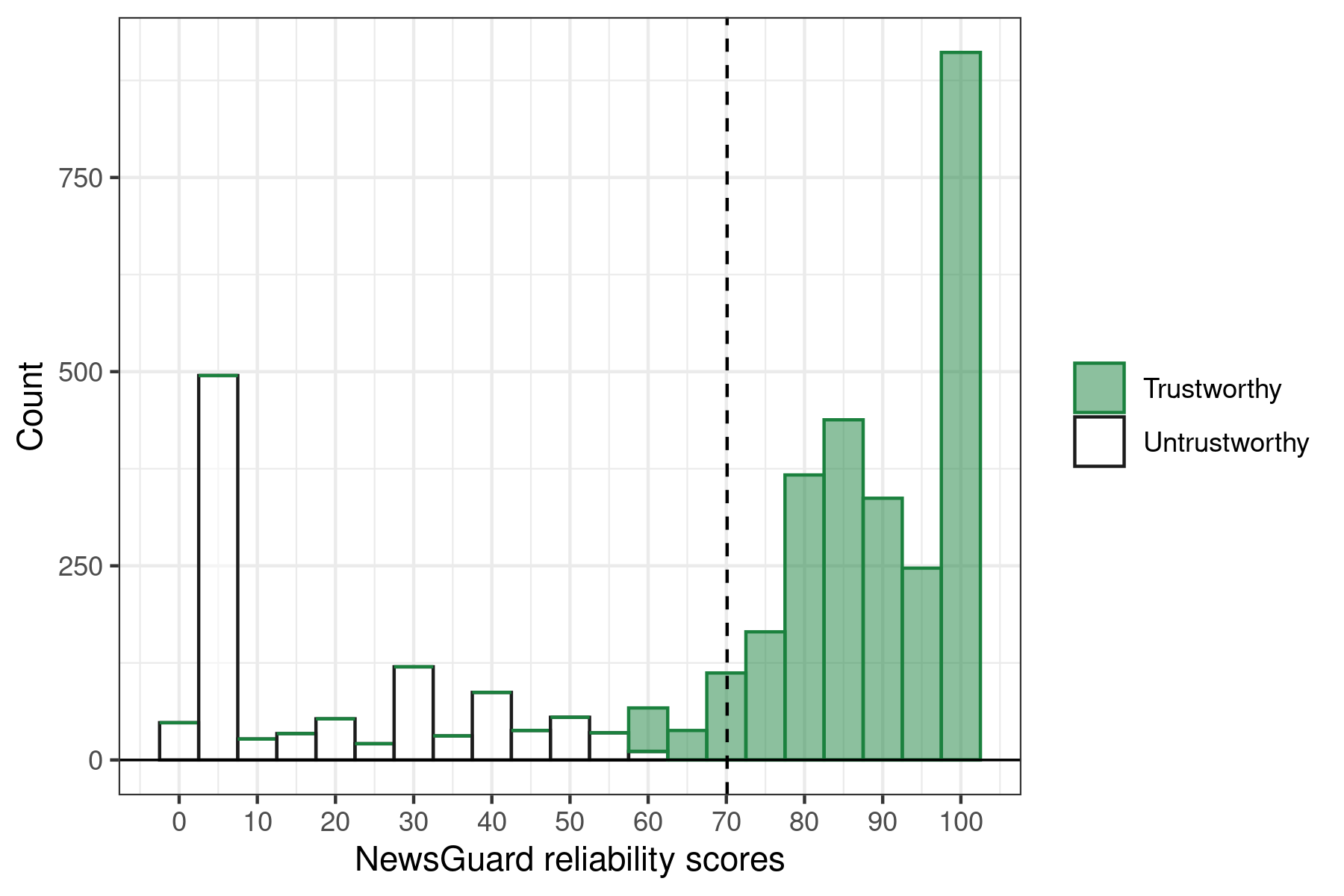}\\
\caption{Distribution of NewsGuard scores ($N = 3{,}726$) by trustworthiness rating. Domains that score below 60 points (i.e.,~untrustworthy) on the rubric used by NewsGuard~\cite{newsguard} are shown in white. Those that score 60 or above are shown in green. The bin width is 5; the bin containing score 60 also includes a few domains with lower scores. The dashed line indicates the average score in the data.}
\label{fig:ng_distribution}
\end{figure}

Our analysis combines two sources of data. The first is the NewsGuard News Website Reliability Index~\cite{newsguard}, a list of web domain reliability ratings compiled by a team of professional journalists and news editors. The data that we licensed for research purposes includes scores of 3,765 web domains on a 100-point scale based on a number of journalistic criteria such as editorial responsibility, accountability, and financial transparency.\footnote{These data were current as of November 12, 2019 and do not reflect subsequent updates; see \hyperref[sec:datastmt]{Code and Data Availability} for more information.} NewsGuard categorizes web domains into four main groups: ``Green'' domains, which have a score of 60 or more points and are considered reliable; ``Red'' domains, which score less than 60 points and are considered unreliable;
``Satire'' domains, which should not be regarded as news sources regardless of their score; and ``Platform'' domains like Facebook or YouTube that primarily host content generated by users. 
The mean reliability score for domains in the data is 69.6; the distribution of scores is shown in Fig.~\ref{fig:ng_distribution}.

The second data source is the YouGov Pulse panel, a sample of U.S.-based Internet users whose web traffic was collected in anonymized form with their prior consent. This traffic data was collected during seven periods between October 2016 and March 2019 (see Table~\ref{tab:pulse_resp}). A total of \samplesize{} participants provided data. Overall, this group is diverse and resembles the U.S. population on key demographic and political dimensions (47.9\% male, 29.0\% with a four-year college degree, 67.9\% white, median age of 55, 37.8\% identifying as Democrats, and 26.3\% identifying as Republicans; see Table ~\ref{tab:pulse_resp} for details by sample collection period). Note that, to be eligible for the study, participants in the YouGov Pulse panel had to be 18+ years of age, so the reported dimensions should be interpreted as being conditional on this extra eligibility criterion.

We perform a number of pre-processing steps on this data. We combine all waves into a single sample. We pool web traffic for each domain that received thirty or more unique visitors. Finally, we use the self-reported partisanship of the visitors (on a seven-point scale from an online survey) to estimate mean audience partisanship and audience partisan diversity, which we estimate using different measures described next and evaluated in the Supplementary Materials.

\begin{table*}
    \begin{threeparttable}
    \footnotesize
    \centering
    \caption{YouGov Pulse respondent data summary}
    \label{tab:pulse_resp}
    \begin{footnotesize}
    \begin{tabular}{c c c c c c c c c c}
        \toprule
\textbf{Dates} & \textbf{Sample} & \textbf{Male} & \textbf{College} & \textbf{White} & \textbf{Age} & \textbf{Dem.} & \textbf{GOP} & \textbf{Domains} & \textbf{Pageviews}\\ 
        \midrule
        10/7--11/14/16 & 3,251 & 47.6\% & 29.2\% & 68.2\% & 58 & 37.2\% & 26.1\% & 158,706  & 26,715,631 \\
        10/25--11/21/17 & 2,100 & 47.6\% & 27.2\% & 69.1\% & 45 & 34.6\% & 25.0\% & 104,513 & 14,247,987 \\
        6/11--7/31/18 & 1,718 & 48.1\% & 30.2\% & 64.9\% & 54 & 38.4\% & 27.4\%  & 108,953 & 15,212,281 \\
        7/12--8/2/18 & 2,000 & 48.8\% & 29.3\% & 65.2\% & 57 & 38.4\% & 26.0\%  & 74,469 & 9,395,659 \\
        10/5--11/5/18 & 3,332 & 48.3\% & 29.0\% & 64.6\% & 55 & 39.5\% & 26.5\%  & 98,850 & 19,288,382 \\
        11/12/18--1/16/19 & 4,907 & 48.7\% & 28.8\% & 64.1\% & 50 & 36.4\% & 26.8\%  & 117,510 & 21,093,638 \\
        1/24--3/11/19 & 2,000 & 48.4\% & 29.6\% & 65.2\% & 55 & 34.9\% & 28.0\%  & 113,700 & 27,482,462 \\
        \bottomrule
    \end{tabular}
        \end{footnotesize}
    \begin{tablenotes}
        \item \small Note: The participants for each data collection period were different. Some participants took part in multiple waves but overlap was small.
    \end{tablenotes}
    \end{threeparttable}
\end{table*}

\subsection{Definition of audience partisan diversity}
\label{sec:ad}

To measure audience partisan diversity, first define $N_j$ as the count of participants who visited a web domain and reported their political affiliation to be equal to $j$ for $j=1,\ldots,7$ (where 1 $=$ strong Democrat and 7 $=$ strong Republican). The total number of participants who visited the domain is thus $N = \sum_j N_j$, and the fraction of participants with a partisanship value of $j$ is $p_j = N_j / N$. Denote the partisanship of the $i$-th individual as $s_i$. 
We calculate the following metrics to measure audience partisan diversity:
\begin{description} 
\item[Variance] $\sigma^2 = N^{-1}\sum(s_i - \overline{s})^2$, where $\overline{s}$ is average partisanship;
\item[Shannon's entropy] $S = - \sum p(j) \log p(j)$, where $p(j)$ is estimated in the following three different ways: (\emph{i}) $p(j) = p_j$ (maximum likelihood); (\emph{ii}) $p(j) = \frac{N_j + \alpha}{N + 7\alpha }$ (mean of the posterior distribution of Dirichlet prior with $\alpha= 1$); and (\emph{iii}) the method of \citeauthor{nemenman2001entropy}~\citep{nemenman2001entropy}, which uses a mixture of Dirichlet priors (NSB prior). 
\item[Complementary Maximum Probability] $1 - \max_j \left\{p_j\right\}$;
\item[Complementary Gini] $1 - G$ where $G$ is the Gini coefficient of the count distribution $\left\{N_j\right\}_{j=1\ldots 7}$.
\end{description}

The above metrics all capture the idea that the partisan diversity of the audience of a web domain should be reflected in the distribution of its traffic across different partisan groups. Each weighs the contribution of each individual person who visits the domain equally; they can thus be regarded as user-level measures of audience partisan diversity. 
However, the volume and content of web browsing activity is highly heterogeneous across internet users~\citep{montgomery2001identifying, guessnd2}, with different users recording different numbers of pageviews to the same website. To account for this imbalance, we also compute the pageview-level, weighted variants of the above audience partisan diversity metrics where, instead of treating all visitors equally, each individual visitor is weighted by the number of pageviews they made to any given domain. 

As a robustness check, we compare the strength of association of each of these metrics to news reliability in the Supplementary Materials. We find that all variants correlate with news reliability, but the relationship is strongest for variance. 

\subsection{Incorporating audience partisan diversity into collaborative filtering recommendations}\label{sec:cf+d}

In general, a recommendation algorithm takes a set of users $\mathcal U$ and a set of items $\mathcal D$ and learns a function $f: \mathcal U \times \mathcal D \rightarrow \mathbb{R}$ that assigns a real value to each user-item pair $\left(u,d\right)$ representing the interest of user $u$ in item $d$. This value denotes the estimated rating that user $u$ will give to item $d$. In the context of the present study, $\mathcal D$ is a set of news sources identified by their web domains (e.g., \texttt{nytimes.com}, \texttt{wsj.com}), so from now on we will refer to $d\in\mathcal D$ interchangeably as either a web domain or a generic item.
 
Collaborative filtering is a classic recommendation algorithm in which some ratings are provided as input and unknown ratings are predicted based on those known input ratings. In particular, the \emph{user-based} CF algorithm, which we employ here, seeks to provide the best recommendations for users by learning from others with similar preferences. CF therefore requires a user-domain matrix where each entry is either known or needs to be predicted by the algorithm. Once the ratings are predicted, the algorithm creates a ranked list of domains for each user that are sorted in descending order by their predicted ratings.
 
To test the standard CF algorithm and our modified CF+D algorithm, we first construct a user-domain matrix $V$ from the YouGov Pulse panel. The YouGov Pulse dataset does not provide user ratings of domains, so we instead count the number of times $\pi_{u,d}\in \mathbb Z^+$ a user $u$ has visited a domain $d$ (i.e.,~pageviews) and use this variable as a proxy~\cite{10.1145/3130332.3130334}. Because this quantity is known to follow a very skewed distribution, we compute the rating as the TF-IDF of the pageview counts:
\begin{linenomath*}
\begin{equation}
 \label{eq:rating}
     v_{u,d} = \frac{\pi_{u,d}}{\sum_{h} \pi_{u,h}}\log\left(\frac{\pi}{\sum_u \pi_{u,d}}\right)
\end{equation}
\end{linenomath*}
where $\pi = \sum_u\sum_d \pi_{u,d}$ is the total number of visits. 
Note that if a user has never visited a particular domain, then $v_{u,d} = 0$. Therefore, if we arrange all the ratings into a user-domain matrix $V\in\mathbb R^{|\mathcal U|\times|\mathcal D|}$, such that $(V)_{u,d} = v_{u,d}$, we will obtain a sparse matrix. The goal of any recommendation task is to complete the user-domain matrix by predicting the missing ratings, which in turn allows us to recommend new web domains to users that may not have seen them. In this case, however, we lack data on completely unseen domains. To test the validity of our methods, we therefore follow the customary practice in machine learning of setting aside some data to be used purely for testing (see Methods~\ref{sec:supervised}). 

Having defined $V$, the next step of the algorithm is to estimate the similarity between each pair of users. To do so, we use either the Pearson correlation coefficient or the Kendall rank correlation of their \emph{user vectors}; i.e.,~their corresponding row vectors in $V$ (i.e., zeroes included). For example, if $\tau(\cdot, \cdot)\in \left[-1,1\right]$ denotes the Kendall rank correlation coefficient between two sets of observations, then the corresponding coefficient of similarity between $u\in\mathcal U$ and $u'\in\mathcal U$ can be defined as:
\begin{linenomath*}
\begin{equation}
\mathrm{sim}(u, u') = \frac{\tau(V_u, V_{u'}) + 1}{2}
\label{eq:sim}
\end{equation}
\end{linenomath*}
where $V_u,V_{u'} \in\mathbb R^{1\times|\mathcal U|}$ are the row vectors of $u$ and $u'$, respectively. A similar definition can be used for Pearson's correlation coefficient in place of $\tau$. 

These similarity coefficients are in turn used to calculate the predicted ratings. In the standard user-based CF, the predicted rating of a user $u$ for a domain $d$ is calculated as:
\begin{linenomath*}
\begin{equation}
\label{eq:cf}
    \hat{v}^{\rm CF}_{u,d} = \bar{v}_u + \frac{\sum_{u' \in N_{u_d}}\mathrm{sim}(u,u')(v_{u',d}-\bar{v}_{u'})}{\sum_{u' \in N_{u_d}}\mathrm{sim}(u,u')}
\end{equation}
\end{linenomath*}
where $N_{u_d} \subseteq \mathcal U$ is the set of the $n=10$ most similar users to $u$ who have also rated $d$ (i.e.,~the neighbors of $u$), $v_{u',d}$ is the observed rating (computed with Eq.~\ref{eq:rating}) that neighboring user $u'$ has given to domain $d$, $\bar{v}_u$ and $\bar{v}_{u'}$ are the average ratings of $u$ and $u'$ across all domains they visited, respectively, and $\mathrm{sim}(u,u')$ is the similarity coefficient (computed with Eq.~\ref{eq:sim}) between users $u$ and $u'$ based on either the Pearson or the Kendall correlation coefficient. 

Having defined the standard CF in Eq.~\ref{eq:cf}, we now define our variant CF+D, which incorporates audience partisan diversity of domain $d\in\mathcal D$ as a re-ranking signal in the following way:
\begin{linenomath*}
\begin{equation}
    \label{eq:cf+d}
    \hat{v}^{\rm CF+D}_{u,d} =  \hat{v}^{\rm CF}_{u,d}+ g\left(\delta_d\right)
\end{equation}
\end{linenomath*}
where $g\left(\delta_d\right)$ is the re-ranking term of domain $d$, obtained by plugging the audience partisan diversity $\delta_d$ (for example, we use the variance of the distribution of self-reported partisan slants of its visitors, $\delta_d = \sigma^2_d$) into a standard logistic function:
\begin{linenomath*}
\begin{equation}
    \label{eq:logistic}
    g(\delta) = \frac{a}{1 + \exp\big(-\left(\delta - t\right)\,/\,\psi\big)}.
\end{equation}
\end{linenomath*}
In Eq.~\ref{eq:logistic}, parameters $a$, $\psi$, and $t$ generalize the upper asymptote, inverse growth rate, and location of the standard logistic function, respectively. For the results reported in this study we empirically estimate the location as $t = \bar{\delta}$, the average audience partisan diversity across all domains, which corresponds to the value of $\bar{\delta} = 4.25$ since we measure diversity as the variance of the distribution of self-reported partisan slants. For the remaining parameters, we choose $a = 1$, $\psi = 1$. As a robustness check, we re-ran all analyses with a larger value of $a$ and obtained qualitatively similar results (available upon reasonable request).

\subsection{Supervised learning evaluation workflow}
\label{sec:supervised}

To evaluate both recommendation algorithms, we follow a standard supervised learning workflow. We use precision and root mean squared error (RMSE), two standard metrics used to measure the relevance and accuracy of predicted ratings in supervised learning settings. We define these two metrics elsewhere (see Methods~\ref{sec:eval}). Here, we instead describe the workflow we followed to evaluate the recommendation methods. Since our approach is based on supervision, we need to designate some of the user ratings (i.e., the number of visits to each domain, which are computed using Eq.~\ref{eq:rating}) as ground truth to compute performance metrics. 

For each user, we randomly split the domains they visited into a training set (70\%) and a testing set (30\%). This splitting varies by user: the same domain could be included in the training set of a user and in the testing set of another. Then, given any two users, their training set ratings are used to compute user-user similarities using Eq.~\ref{eq:sim} (which is based on Kendall's rank correlation coefficient; a similar formula can be defined using Pearson's correlation).
If, in computing user-user similarities with Eq.~\ref{eq:sim}, a domain is present for a user but not for the other, then the latter rating is assumed to be zero regardless of whether the domain is present in testing or not. This assumption, which follows standard practice in collaborative filtering algorithm, ensures that there is no leaking of information between the test and training sets.

Finally, using either Eq.~\ref{eq:cf} or Eq.~\ref{eq:cf+d}, we predict ratings for domains in the test set and compare them with the TF-IDF of the actual visit counts in the data. 

\subsection{Recommendation based on global popularity}\label{sec:global_popularity}

We also generate ranked lists for users based on global domain popularity (user-level) as an additional baseline recommendation technique. All the domains are initially assigned a rank (global popularity rank) according to their user-level popularity, which is calculated from the training set views. Then, the domains in the test set of each user are ranked according to their global popularity ranks to generate the recommendations. This method does not include any personalization as the rank of a domain for a particular user does not depend on other similar users but depends on the whole population. In particular, if two users share the same two domains in testing, their relative ranking is preserved, even if the two users visited different domains in training.

\subsection{Trustworthiness metrics}
\label{sec:trust}

In addition to standard metrics of accuracy (precision and RMSE; see Methods~\ref{sec:eval}), we define a new metric called \emph{trustworthiness} to measure the news reliability of the recommended domains. It is calculated using NewsGuard scores in two ways: either using the numerical scores or the set of binary indicators for whether a site meets or exceeds the threshold score of 60 defined by NewsGuard as indicating that a site is generally trustworthy~\cite{newsguard}. Let $d_1,d_2,\dots,d_k$ be a ranked list of domains. Using numerical scores, the trustworthiness is the average:
\begin{linenomath*}
\begin{equation}
    \label{eq:tw}
    \frac{1}{k}\sum\limits_{r=1}^{k}Q(d_{r})
\end{equation}
\end{linenomath*}
where $Q(d) \in \left[0, 100\right]$ denotes the NewsGuard reliability score of $d\in\mathcal D$.

If instead we use the binary indicator of trustworthiness provided by Newsguard, then the trustworthiness of domains in a list is defined as the fraction of domains that meet or exceed the threshold score. 
Note that, unlike precision and RMSE, the trustworthiness of a list of recommendations does not use information on the actual ratings $v_{u,d}$. Instead, using Eq.~\ref{eq:tw}, we compute the trustworthiness of the domains in the test set ranked in decreasing order of user visits $v_{u,d}$. We then compare the trustworthiness of the rankings obtained with either CF or CF+D against the trustworthiness of this baseline.

\subsection{Accuracy metrics}
\label{sec:eval}

Given a user $u$, let us consider a set $\mathcal D$ of web domains for which $|\mathcal D| = D$. For each domain $d \in \mathcal D$, we have three pieces of information: the two predicted ratings $\hat{v}^{\mathrm{CF}}_{u,d}$ and $\hat{v}^{\mathrm{CF+D}}_{u,d}$ produced by CF and CF+D and the actual rating $v_{u, d}$ (defined elsewhere; see Methods~\ref{sec:cf+d}). In the following, we omit the subscript $u$ of the user, which is fixed throughout, and the CF/CF+D superscript unless it is not obvious from  context.

Let us consider a given recommendation method (either CF or CF+D) and denote with $r(d)$ (respectively, $r'(d)$) the rank of $d$ when the domains are sorted by decreasing order of recommendation and actual ratings, respectively. Given a recommendation list length $0 < k \le D$, let us define the set of predicted domains as:
\begin{linenomath*}
\[
P_k = \{d \in \mathcal D : r(d) \le k \}
\]
\end{linenomath*}
and the set of actual domains as:
\begin{linenomath*}
\[
A_k = \{d \in \mathcal D : r'(d) \le k \}.
\]
\end{linenomath*}
Then the \emph{precision} for a given value of $k$ is given by the fraction of correctly predicted domains:
\begin{linenomath*}
\[
\mathrm{Precision} = \frac{|P_k \cap A_k|}{|P_k|}.
\]
\end{linenomath*}
Similarly, the \emph{root mean squared error} for a given value of $k$ between the two ranked lists of ratings is computed as:
\begin{linenomath*}
\[
\mathrm{RMSE} = \sqrt{\frac{1}{k} \sum_{r = 1}^{k} \left(\hat{v}_{\rho(r)} - v_{\rho'(r)}\right) ^ 2 }
\]
\end{linenomath*}
where $\rho:[D]\mapsto\mathcal D$ (respectively $\rho'$) is the inverse function of $r(\cdot)$ (respectively, $r'(\cdot)$); that is, the function that maps ranks back to their domain by the recommendation method (respectively, by actual visits). Note that, in the summation, $\rho(r)$ and $\rho'(r)$ do not generally refer to the same web domain: the averaging is over the two ranked lists of ratings, not over the set of domains in common between the two lists. 

\subsection{Discounting via ranking}
\label{sec:rankingmodel}

To measure the effect of CF+D on the trustworthiness of rankings, we must select a particular list length $k$. Although Fig.~\ref{fig:trust} shows improvements for all values of $k$, one potential problem when stratifying on different groups of users is that the results could depend on the particular choice of $k$. To avoid dependence on $k$, we consider a probabilistic model of a hypothetical user visiting web domains from a ranked list of recommendations 
and define overall trustworthiness as the expected value of the trustworthiness of domains selected from that list (i.e., discounted by probability of selection). 

Let us consider a universe of domains $\mathcal D$ as the set of items to rank. Inspired by prior approaches on stochastic processes based on ranking~\cite{fortunato2006scale}, we consider a discounting method that posits that the probability of selecting domain $d\in\mathcal D$ from a given ranked recommendation list decays as a power law of its rank in the list:
\begin{linenomath*}
\begin{equation}\label{eq:ranking}
    \Pr\left\{X = d\right\} = \frac{r_d^{-\alpha}}{\sum_{h}r_h^{-\alpha}}
\end{equation}
\end{linenomath*}
where $X\in\mathcal D$ is a random variable denoting the probabilistic outcome of the selection from the ranked list, $r_d \in \mathbb N$ is the rank of a generic $d\in\mathcal D$, and $\alpha \ge 0$ is the exponent of power-law decay (when $\alpha = 0$, all domains are equally likely; when $\alpha > 0$, top-ranked domains are more likely to be selected).

This procedure allows us to compute, for any given user, the effect of a recommendation method (either CF or CF+D) simply as the difference between its expected trustworthiness and the trustworthiness of the ranking obtained by sorting the domains visited by the user in decreasing order of pageviews (see Eq.~\ref{eq:change}).

In practice, to compute Eq.~\ref{eq:change}, let $d_1,d_2,\dots,d_k$ and $d'_1,d'_2,\dots,d'_k$ be two ranked lists of domains, $d_r,d'_r\in\mathcal D~\forall r=1,\ldots,k$, generated by a recommendation algorithm and by actual user pageviews, respectively, and let us denote with $Q(d)$ the NewsGuard reliability score of $d\in\mathcal D$ (see Methods~\ref{sec:trust}). 
Recall that Eq.~\ref{eq:ranking} specifies the probability of selecting a given domain $d\in\mathcal D$ from a particular ranked list as a function of its rank. Even though any pair of equally-ranked domains will be different across these two lists (that is, $d_r \neq d_r'$ in general), their probability will be the same because Eq.~\ref{eq:ranking} only depends on $r$. We can thus calculate the expected improvement in trustworthiness as:
\begin{linenomath*}
\begin{equation}
    \Delta Q 
    = \sum_{r=1}^{k} P(r) \left(Q\left(d_{r}\right) - Q\left(d'_{r}\right)\right)
    \label{eq:expected}
\end{equation}
\end{linenomath*}
where $P(r)$ is the probability of selecting a domain with rank $r$ from Eq.~\eqref{eq:ranking}, which we computed setting $\alpha = 1$. 

\subsection{Stratification analysis}
\label{sec:stratification}

Recall that we use the self-reported partisanship of respondents in the YouGov Pulse panel as the basis for our diversity signal (see Methods~\ref{sec:ad}). To avoid the circular reasoning in stratifying on the same source of data, Fig.~\ref{fig:tw_vs}(a) and Fig.~\ref{fig:tw_vs}(c) group these users according to the slant of their actual news consumption, which may not necessarily reflect their self-reported partisanship (e.g., a self-reported Democrat might access mostly conservative-leaning websites). We determined this latter metric using an external classification originally proposed by \citeauthor{BakshyAdamic}~\cite{BakshyAdamic}, who estimated the slant of 500 web domains focused on hard news topics. In practice, \citeauthor{BakshyAdamic} based their classification on how hard news from those domains were shared on Facebook by users who self-identified as liberal or conservative in their profile. For almost all domains, \citeauthor{BakshyAdamic} reported a value $s\in [-1,1]$ with a value of $s = +1$ for domains that are shared almost exclusively by conservatives, and a value of $s = -1$ for those shared almost exclusively by liberals. (These values could technically vary over $[-2,2]$ but only 1\% of domains fell outside $[-1,1]$ using the measurement approach described by \citeauthor{BakshyAdamic}~\cite{BakshyAdamic}.)

In Fig.~\ref{fig:tw_vs}(c), respondents are grouped according to the absolute slant $\left| s \right|$ of the visited domains where a value of $\left|s \right| = 0$ denotes domains with a perfectly centrist slant and a value of $\left|s \right| = 1$ indicates domains with extreme liberal or conservative slants (i.e., they are almost exclusively shared by one group and not the other).

\section*{Author Contributions}

All authors designed the research. S.B. and S.Y. performed data analysis. All authors wrote, reviewed, and approved the manuscript.

\section*{Ethics Statement}

This study was reviewed by the IRB under protocols \#HUM00161944 (University of Michigan) and \#STUDY000433 (University of South Florida).

\section*{Code and Data Availability}
\label{sec:datastmt}

Data necessary to reproduce the findings in the manuscript are available, in aggregated and anonymized format, at \url{https://github.com/glciampaglia/InfoDiversity/} along with the associated source code. To reproduce the findings in the Supplementary Materials, additional code and data are available upon reasonable request. 

The raw data that support the findings of this study are available from NewsGuard Technology, Inc. but restrictions apply to the availability of these data, which were used under license for the current study and thus cannot be made publicly available. However, data are available from the authors upon reasonable request subject to licensing from NewsGuard. The data used in this study were current as of November 12, 2019 and do not reflect NewsGuard's regular updates of the data.

\section*{Acknowledgements}

We thank NewsGuard for licensing the data and acknowledge Andrew Guess and Jason Reifler, Nyhan's coauthors on the research project that generated the web traffic data used in this study. We are also grateful to organizers, chairs and participants of the News Quality in the Platform Era workshop (organized by the Social Science Research Council), especially Regina Lawrence, Philip Michael Napoli, Kevin Munger, Johanna Dunaway, Connie Moon Sehat and Jieun Shin, for their helpful comments. This work was supported in part by the National Science Foundation under a collaborative award (NSF Grant No.~1915833, 1949077). Any opinions, findings, and conclusions or recommendations expressed in this material are those of the authors and do not necessarily reflect the views of the National Science Foundation.


\clearpage
\newcommand{\beginsupplement}{%
        \setcounter{section}{0}
        \renewcommand{\thesection}{S\arabic{section}}%
        \setcounter{table}{0}
        \renewcommand{\thetable}{S\arabic{table}}%
        \setcounter{figure}{0}
        \renewcommand{\thefigure}{S\arabic{figure}}%
     }

\beginsupplement{}

\begin{minipage}{.95\textwidth}
\begin{center}
    \LARGE\textbf{Supplementary Materials for Bhadani \emph{et al.}, ``\emph{\thetitle{}}''}
\end{center}
\end{minipage}
\vspace{2em}

\section{Alternative Definitions of Audience Diversity} \label{sec:alt}

We repeat the analysis of Fig.~\ref{fig:average_slant_vs_variance2} for all diversity metrics (see Methods~\ref{sec:ad}) and summarize the results in Table~\ref{tab:metrics}. For each metric, we estimate the degree of linear association with news quality using the Pearson correlation coefficient. We also report the $R^2$ coefficient of determination and the two-sided $p$-value of the F-statistic as a measure of significance of the fit. And finally, we show the partial correlation coefficient by controlling the mean partisanship and the extremity of domains. Each metric is positively correlated with quality at the user level, but we find that the relationship is strongest for variance of audience partisanship. At the pageview level, however, the association disappears for all metrics but variance, which still produces a modest correlation. 

\begin{table}[h!]
\footnotesize
\caption{Relationship between audience partisan diversity and news quality.}
\label{tab:metrics}
\begin{tabular}{lccccccc}\toprule
\textbf{Diversity metric} & {\bf Correlation} & {R$^2$} & {$p$-value} & \multicolumn{2}{c}{Partial Correlation (Mean)} & \multicolumn{2}{c}{Partial Correlation (Extremity)}
\\\cmidrule(lr){5-6}\cmidrule(lr){7-8}
 & & & & {\bf Correlation}  & {$p$-value}    & {\bf Correlation}  & {$p$-value}\\\midrule
\multicolumn{8}{c}{\sc user level} \\
\midrule
Variance        & 0.32  & 0.10 & $<0.01$ & 0.38 & $<0.01$ & 0.26 & $<0.01$ \\
Entropy (Dir.)  & 0.21  & 0.04 & $<0.01$ & 0.39 & $<0.01$ & 0.31 & $<0.01$ \\
Entropy (ML)    & 0.20  & 0.04 & $<0.01$ & 0.34 & $<0.01$ & 0.24 & $<0.01$\\
Entropy (NSB)   & 0.22  & 0.05 & $<0.01$ & 0.2 & $<0.01$ & 0.14 & $<0.01$\\
Compl. Max. Prob.    & -0.04 & 0.00 & 0.14 & 0.26 & $<0.01$ & 0.14 & $<0.01$\\
Compl. Gini       & 0.14  & 0.02 & $<0.01$ & 0.26 & $<0.01$ & 0.21 & $<0.01$\\
\midrule
\multicolumn{8}{c}{\sc pageview level} \\
\midrule
Variance        & 0.14 & 0.02 & $<0.01$ & 0.22 & $<0.01$ & 0.15 & $<0.01$ \\
Entropy (Dir.)  & 0.03  & 0.00 & 0.24 & 0.044 & 0.07 & 0.04 & 0.09 \\
Entropy (ML)    & 0.03  & 0.00 & 0.19 & 0.046 & 0.057 & 0.042 & 0.078\\
Entropy (NSB)   & 0.03  & 0.00 & 0.18 & 0.048 & 0.05 & 0.044 & 0.07\\
Compl. Max. Prob.    & 0.004 & 0.00 & 0.86 & 0.03 & 0.22 & 0.019 & 0.42\\
Compl. Gini       & -0.001  & 0.00 & 0.97 & 0.019 & 0.43 & 0.017 & 0.46\\\bottomrule
\end{tabular}
\end{table}

\clearpage

\section{Regression of NewsGuard scores on website audience variance} \label{sec:regression}

In Fig.~\ref{fig:average_slant_vs_variance2} in the main text we show the relationship between NewsGuard reliability scores of news domains and audience partisan diversity, via linear regression. In Tables~\ref{tab:userlevel}--\ref{tab:pageviewlevel} we report the associated summary tables for both the user and pageview level, respectively. To ensure the regression coefficients and associated errors can be comparable across datasets, we standardize all independent variables prior to fitting the models to the data.

\begin{table}[!htbp] 
  \centering 
  \caption{Relationship between NewsGuard scores and user-level partisan audience diversity} 
  \label{tab:userlevel} 
\begin{tabular}{@{\extracolsep{5pt}}lD{.}{.}{-3} D{.}{.}{-3} D{.}{.}{-3} } 
\toprule 
 & \multicolumn{1}{c}{\bf All} & \multicolumn{1}{c}{\bf Republican websites} & \multicolumn{1}{c}{\bf Democratic websites} \\ 
\midrule 
 User-level variance in partisanship & 6.6701^{***} & 13.051^{***} & 2.7671^{***} \\ 
  & (0.579) & (1.589) & (0.535) \\ 
 Constant & 87.1203^{***} & 76.554^{***} & 89.2416^{***} \\ 
  & (0.798) & (2.213) & (0.705) \\ 
\midrule
Observations & \multicolumn{1}{c}{1,020} & \multicolumn{1}{c}{237} & \multicolumn{1}{c}{783} \\ 
R$^{2}$ & \multicolumn{1}{c}{0.115} & \multicolumn{1}{c}{0.223} & \multicolumn{1}{c}{0.033} \\ 
\bottomrule \\[-1.8ex] 
\textit{Note:} Standard errors in parentheses.  
& \multicolumn{3}{r}{$^{*}$p$<$0.1; $^{**}$p$<$0.05; $^{***}$p$<$0.01} \\ 
\end{tabular} 
\end{table}

\begin{table}[!htbp] \centering 
  \caption{Relationship between NewsGuard scores and pageview-level partisan audience diversity} 
  \label{tab:pageviewlevel} 
\begin{tabular}{@{\extracolsep{5pt}}lD{.}{.}{-3} D{.}{.}{-3} D{.}{.}{-3} } 
\toprule 
 & \multicolumn{1}{c}{\bf All} & \multicolumn{1}{c}{\bf Republican websites} & \multicolumn{1}{c}{\bf Democratic websites} \\ 
\midrule 
 Pageview-level variance in partisanship & 3.9194^{***} & 9.614^{***} & 0.6766 \\ 
  & (0.709) & (2.195) & (0.608) \\ 
 Constant & 85.015^{***} & 73.111^{***} &  87.934^{***} \\ 
  & (0.848) & (2.527) & (0.720) \\ 
\midrule 
Observations & \multicolumn{1}{c}{1,020} & \multicolumn{1}{c}{237} & \multicolumn{1}{c}{783} \\ 
R$^{2}$ & \multicolumn{1}{c}{0.029} & \multicolumn{1}{c}{0.075} & \multicolumn{1}{c}{0.002} \\ 
\bottomrule \\[-1.8ex] 
\textit{Note:} Standard errors in parentheses.  
& \multicolumn{3}{r}{$^{*}$p$<$0.1; $^{**}$p$<$0.05; $^{***}$p$<$0.01} \\ 
\end{tabular} 
\end{table}

\clearpage

\section{Correlations between domain popularity and audience diversity}

In Table~\ref{tab:cor_pop_var} we show the Pearson correlation coefficients between the popularity of a domain and its diversity. We operationalize the popularity of website as either its audience size (i.e., number of unique users) or its traffic (number of pageviews). For our diversity measures, we rely on the user-level and pageview-level partisanship variance. 

Overall, domain popularity is very weakly correlated with the variance of audience partisanship regardless of how we choose to operationalize each measure. Recall that in our original analysis we show that domain popularity is largely uncorrelated with quality (as proxied by NewsGuard scores). Together, these findings suggest that audience partisan diversity is associated with quality of news independent of the variation caused by domain popularity.

\begin{table}[h!]
    \centering
    \caption{Pearson correlation coefficients between domain diversity and popularity}
    \begin{tabular}{ccc}
      \toprule
      Variance (rows) / Popularity (columns) & $N$ Unique users & $N$ Pageviews \\
      \midrule
      User-level & $0.04$ ($p = 1.39\times 10^{-5}$) & $0.0093$ ($p = 0.31$) \\
      Pageview-level & $0.062$ ($p = 2.1\times 10^{-11}$) & $0.019$ ($p = 0.038$) \\
      \bottomrule
    \end{tabular}
    \label{tab:cor_pop_var}
\end{table}

Furthermore, we estimate multivariate regressions interacting our diversity measures with an indicator of whether the website has a predominantly Democratic or Republican website with the following model: 
\begin{linenomath*}
\begin{align*}
   \text{Reliability} &= \beta_0 + \beta_1 \text{(Diversity measure)} + \beta_2 \text{(Conservative website dummy)} + \\
   & \beta_3 \text{(Logged audience size)} + \beta_4 \text{(Diversity measure} \times \text{Conservative website dummy)}
\end{align*}
\end{linenomath*}

We estimate two separate regression models, using our two operational diversity measures: user-level and pageview-level partisanship variance. At the user level, after controlling for audience size, a one-standard deviation increase in diversity is associated with a 2.91 point increase in NewsGuard reliability for websites with predominantly Democratic audiences, and with a 10.8 point increase for websites with predominantly Republican audiences. Both estimates are statistically significant. 

At the pageview level, we find that a one-standard deviation increase is associated with a 0.68 point increase (statistically indistinguishable from zero) in NewsGuard reliability for Democratic websites, and with a 9.24 point increase for Republican websites. In summary, both methods indicate that our diversity measure is a good predictor of journalistic quality, independent of audience size. This relationship is especially strong for websites with predominantly Republican audiences. 

\clearpage

\section{Robustness Checks}
%

\begin{figure}
     \includegraphics[width=0.495\textwidth]{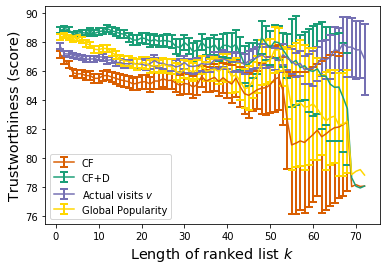}
     \includegraphics[width=0.495\textwidth]{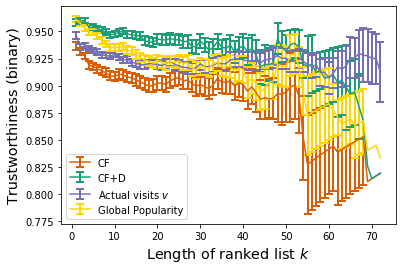}
    \caption{Trustworthiness of recommended domains by length of ranked list $k$, for all values of $k$. Left: Trustworthiness based on scores from NewsGuard~\cite{newsguard}. Right: proportion of domains labeled as `trustworthy,' also by NewsGuard. Actual visits $v$ are normalized using TF-IDF (see Methods~\ref{sec:cf+d}). Each bin represents the average computed on the top-$k$ recommendations for all users in the YouGov panel with $\ge k$ recommendations in their test sets. Bars represent the standard error of the mean. In this figure, both CF and CF+D compute the similarity between users using the Kendall $\tau$ correlation coefficient (see Methods~\ref{sec:cf+d}).}
    \label{fig:trust_kendall_all}
\end{figure}

\begin{figure}
     \includegraphics[width=0.495\textwidth]{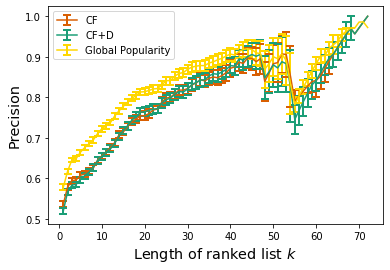}
     \includegraphics[width=0.495\textwidth]{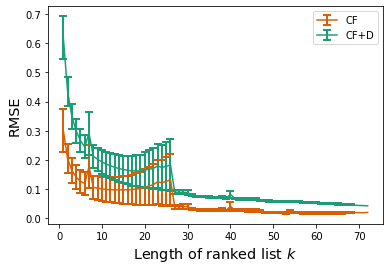}
    \caption{Accuracy of domain recommendations by length of ranked list $k$, for all values of $k$. Left: Precision (proportion of correctly ranked sites) by length of ranked list $k$ (higher is better). Right: RMSE (root mean squared error) of predicted pageviews for top $k$ ranked domains by length of ranked list $k$ (lower is better). Each bin represents the average computed on the top-$k$ recommendations of all users with $\ge k$ recommendations in their test sets. Bars represent the standard error of the mean. In the last bin ($k=73$) precision is 100\% for all users. In this figure, both CF and CF+D compute the similarity between users using the Kendall $\tau$ correlation coefficient (see Methods~\ref{sec:cf+d}).}
    \label{fig:prec_rmse_kendall_all}
\end{figure}

\paragraph{No minimum frequency capping.} 
Figs.~\ref{fig:trust_kendall_all} and~\ref{fig:prec_rmse_kendall_all} are the analogous of Figs.~\ref{fig:trust} and~\ref{fig:precision} from the main text, but unlike the plots in the main text, which capped the range of $k$ to include only bins with a minimum frequency, the plots here show all possible values of $k$. 

\begin{figure}
     \includegraphics[width=0.495\textwidth]{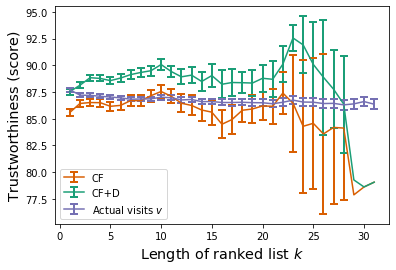}
     \includegraphics[width=0.495\textwidth]{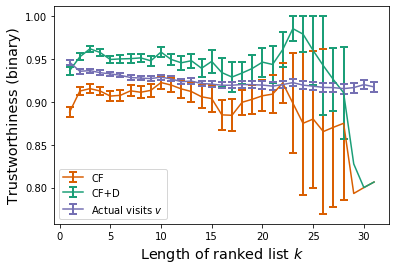}
    \caption{Trustworthiness of recommended domains by length of ranked list $k$, for all values of $k$. Left: Trustworthiness based on scores from NewsGuard~\cite{newsguard}. Right: proportion of domains labeled as `trustworthy,' also by NewsGuard. Actual visits $v$ are normalized using TF-IDF (see Methods~\ref{sec:cf+d}). All results represent averages computed for all users in the YouGov panel. Bars represent the standard error of the mean. In this figure, both CF and CF+D compute the similarity between users using the Pearson correlation coefficient.}
    \label{fig:trust_pearson}
\end{figure}

\begin{figure}
     \includegraphics[width=0.495\textwidth]{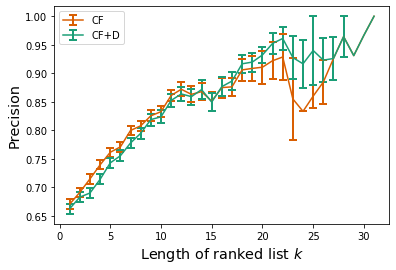}
     \includegraphics[width=0.495\textwidth]{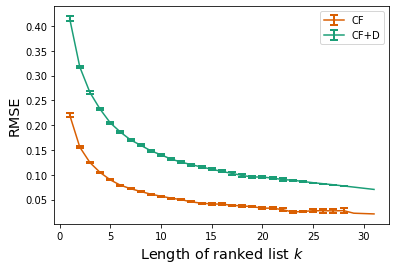}
    \caption{Accuracy of domain recommendations by length of ranked list, for all values of $k$. Left: Precision (proportion of correctly ranked sites) by length of ranked list $k$ (higher is better). Right: RMSE (root mean squared error) of predicted pageviews for top $k$ ranked domains by length of ranked list $k$ (lower is better). Each bin represents the average computed on the top-$k$ recommendations of all users with $\ge k$ recommendations in their test sets. Bars represent the standard error of the mean. In the last bin ($k=30$) precision is 100\% for all users. Bars represent the standard error of the mean. In this figure, both CF and CF+D compute the similarity between users using the Pearson correlation coefficient.}
    \label{fig:prec_rmse_pearson}
\end{figure}

\begin{figure}
    \includegraphics[width=0.495\textwidth]{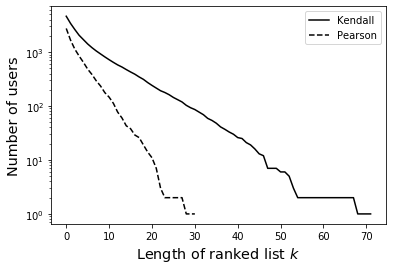}
    \caption{Number of users with $k$ domains in the test set for neighborhoods (the set of the $n=10$ most similar users to a given user) computed using the correlation coefficient of Kendall (solid line) and Pearson (dashed line). In general, Pearson leads to shorter lists of recommendations.}
    \label{fig:size}
\end{figure}

\paragraph{Alternative similarity metric based on rank correlation.}
Figs.~\ref{fig:trust_pearson} and \ref{fig:prec_rmse_pearson} also show the results of analyses analogous to those in Figs.~\ref{fig:trust} and \ref{fig:precision}, but unlike the plots in the main text, which used the Kendall rank correlation coefficient to compute the similarity between users, the plots here show the results obtained using the Pearson correlation coefficient. Moreover, the plots here show all possible values of $k$, without the aforementioned cap. To get a better sense sense of this difference, Fig.~\ref{fig:size} shows the distribution of the number of users as a function of the length of the ranked list $k$. We observe that Pearson tends to produce smaller recommendation lists than Kendall.

\begin{figure}
     \includegraphics[width=0.495\textwidth]{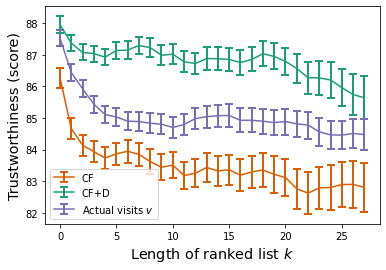}
     \includegraphics[width=0.495\textwidth]{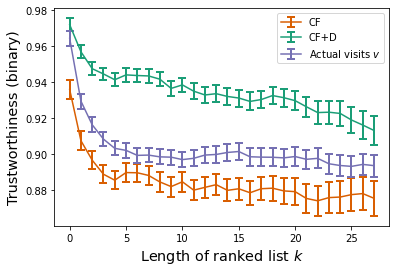}
    \caption{Trustworthiness of recommended domains by length of ranked list $k$ when the data for training and test sets for the first wave of users are split based on time Left: Trustworthiness based on scores from NewsGuard~\cite{newsguard}. Right: proportion of domains labeled as `trustworthy,' also by NewsGuard. Actual visits $v$ are normalized using TF-IDF (see Methods~\ref{sec:cf+d}). Each bin represents the average computed on the top-$k$ recommendations for all users in the YouGov panel with $\ge k$ recommendations in their test sets. Bars represent the standard error of the mean. In this figure, both CF and CF+D compute the similarity between users using the Kendall $\tau$ correlation coefficient (see Methods~\ref{sec:cf+d}).}
    \label{fig:trust_kendall_time}
\end{figure}

\begin{figure}
     \includegraphics[width=0.495\textwidth]{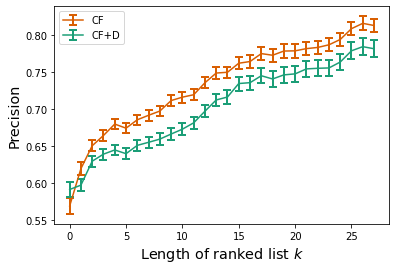}
     \includegraphics[width=0.495\textwidth]{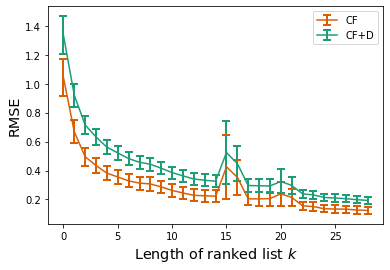}
    \caption{Accuracy of domain recommendations by length of ranked list $k$ when the data for training and test sets for the first wave of users are split based on time. Left: Precision (proportion of correctly ranked sites) by length of ranked list $k$ (higher is better). Right: RMSE (root mean squared error) of predicted pageviews for top $k$ ranked domains by length of ranked list $k$ (lower is better). Each bin represents the average computed on the top-$k$ recommendations of all users with $\ge k$ recommendations in their test sets. Bars represent the standard error of the mean. In the last bin ($k=73$) precision is 100\% for all users. In this figure, both CF and CF+D compute the similarity between users using the Kendall $\tau$ correlation coefficient (see Methods~\ref{sec:cf+d}).}
    \label{fig:prec_rmse_kendall_time}
\end{figure}

\paragraph{Longitudinal analysis.} 
Figs.~\ref{fig:trust_kendall_time} and~\ref{fig:prec_rmse_kendall_time} show the results of an analysis analogous to those in Figs.~\ref{fig:trust} and \ref{fig:precision}, but in which training and testing sets are split longitudinally instead of randomly. In this sense, they represent a true forecasting exercise. Despite a slightly larger loss of precision relative to CF (compare the left panel of Fig.~\ref{fig:precision} in the main text with the left panel of Fig.~\ref{fig:prec_rmse_kendall_time}), our results remain qualitatively consistent with those shown in the main text.
For the prior Figs.~\ref{fig:trust},~\ref{fig:precision},~\ref{fig:trust_kendall_all},~\ref{fig:prec_rmse_kendall_all},~\ref{fig:trust_pearson} and~\ref{fig:prec_rmse_pearson}, the data for each user are randomly split into a training (70\%) and testing set (30\%), so that, for any given user, there is no overlap between the two sets. Note that each user is split independently of the others, so a given domain can appear in the training set of one user and in the testing set of another. 
Instead, in Figs.~\ref{fig:trust_kendall_time} and~\ref{fig:prec_rmse_kendall_time}, the data of traffic that took place before a fixed boundary date (which is identical for all users) form the training set, and those that took place after form the testing set. This means that the same domain can occur in both the training and the testing set. 

Data collection for the YouGov Pulse panel took place in 7 different time periods (see Table~\ref{tab:pulse_resp}), but for simplicity we considered only 3 waves (the first three). Figs.~\ref{fig:trust_kendall_time} and~\ref{fig:prec_rmse_kendall_time} show the analysis performed on the first wave of data collection, which took place between October 7 and November 14, 2016, and we split the data using November 1, 2016 as boundary. We find qualitatively similar results for the second and third waves. (Data available upon reasonable request to the authors.) 

\begin{figure}
     \includegraphics[width=0.495\textwidth]{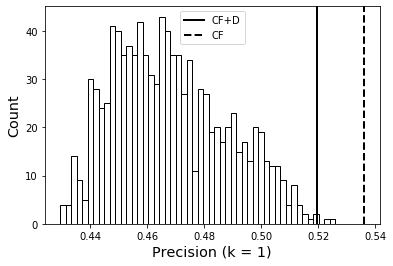}
     \includegraphics[width=0.495\textwidth]{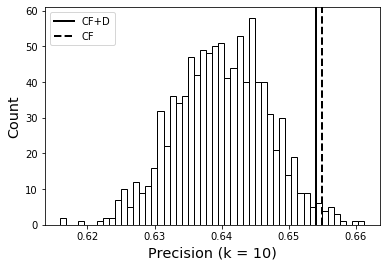}
     \caption{Distribution of precision obtained after re-ranking the domains, by means of re-shuffling the diversity signal values $g(\delta_d)$ from the CF+D ratings calculation (see Eq.~\ref{eq:cf+d} and Eq.~\ref{eq:logistic}). The re-shuffling was repeated $1,000$ times. The two distributions correspond to different values of $k$. The (one-sided) $p$-values are 0.002 ($k$ = 1)
     %
     %
     and 0.021 ($k$ = 10). The two vertical lines correspond to the observed precision values of CF+D (solid) and CF (dashed).} 
     %
     \label{fig:resampling}
\end{figure}

\paragraph{Resampling.} To estimate the significance of the observed drop in precision of CF+D, we simulate the process of re-ranking a list of items. Recommendations are obtained in this context by sorting items by their predicted rating. Since CF+D simply shifts the rating of each item by adding a term that depends on diversity (see Eq.~\ref{eq:cf+d}), we simulate this process by simply shuffling the diversity terms among the items before ranking them. This procedure ensures that we consider only lists obtained by shifting the ratings by the same amount of CF+D. Fig.~\ref{fig:resampling} shows the sampling distribution of the precision of re-rankings of the same magnitude as those of CF+D using this process for $k=1$ and $k=10$. To sample from this distribution, we rank domains using the ratings computed from Eq.~\ref{eq:cf}. We then compute in a separate labeled vector the diversity term $g(\delta_d)$ obtained using the logistic function (Eq.~\ref{eq:logistic}), reshuffle the labels at random, obtaining for each term a new label $d'$, and finally apply the reshuffled term $g(\delta_{d'})$ as in Eq.~\ref{eq:cf+d}. We then re-rank based on the new ratings and compute the precision of the ranked list. This reshuffling is carried out separately for each user with at least $k$ domains in their testing set. The precision is then averaged over all users. This procedure is repeated 1,000 times to obtain the sampling distribution. Finally, we compute a one-tailed $p$-value by finding the proportion of samples that have a precision higher than the observed value for CF+D.

\begin{figure}
    \includegraphics[width=\textwidth]{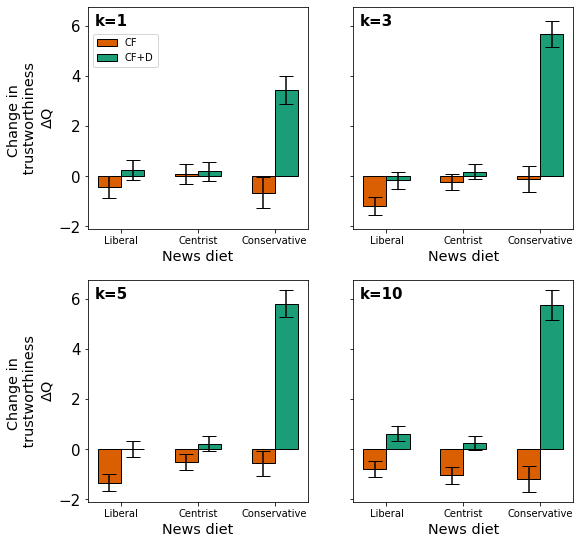}
    \caption{Effect of CF and CF+D versus baseline by ideological slant of visited domains (terciles using scores from \citeauthor{BakshyAdamic}~\cite{BakshyAdamic}) and by length of ranked list $k$. In this and the following plots, bars represent the standard error of the mean. Change in trustworthiness $\Delta Q$ based on scores from NewsGuard~\cite{newsguard}.}
\label{fig:stratification-start}
\end{figure}

\begin{figure}
    \includegraphics[width=\textwidth]{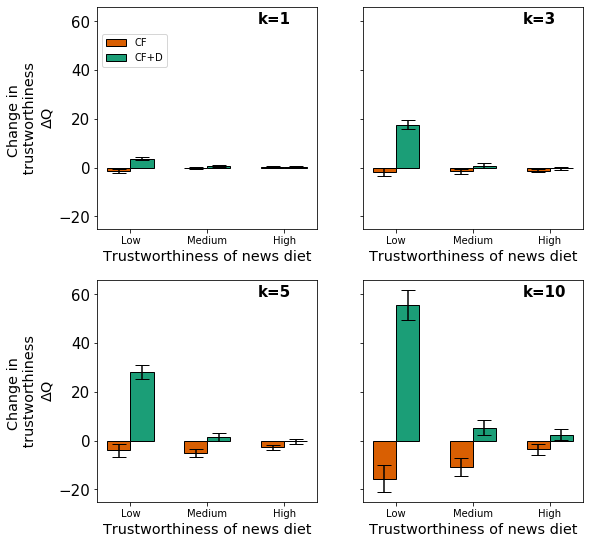}
    \caption{Effect of CF and CF+D versus baseline by baseline trustworthiness of domains visited by users (terciles) and by length of ranked list $k$. Change in trustworthiness $\Delta Q$ based on scores from NewsGuard~\cite{newsguard}.} 
    \label{fig:stratification-end}
\end{figure}

\paragraph{Stratification analysis without discounting.} Fig.~\ref{fig:stratification-start}--\ref{fig:stratification-end} show the results of the stratification analysis without using the discounting model.

\begin{figure}
    \includegraphics[width=\textwidth]{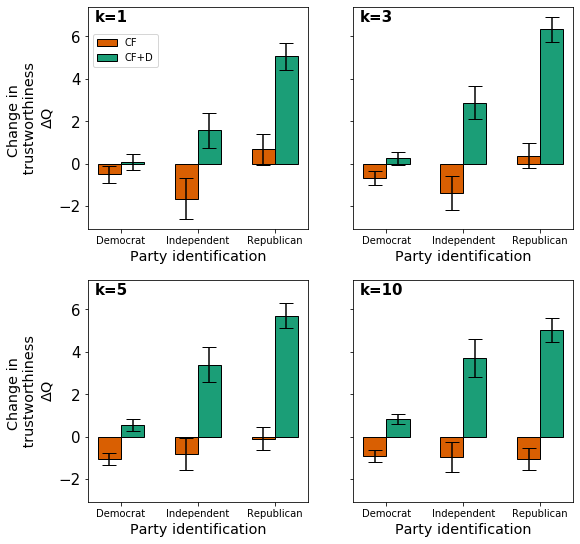}
    \caption{Effect of CF and CF+D versus baseline by self-reported party ID from YouGov Pulse responses as measured on a 7-point scale (1--3: Democrats including people who lean Democrat but do not identify as Democrats, 4: Independents, 5--7: Republicans including people who lean Republican but do not identify as Republicans) and by length of ranked list $k$. Change in trustworthiness $\Delta Q$ based on scores from NewsGuard~\cite{newsguard}.}
\end{figure}

\begin{figure}
    \includegraphics[width=\textwidth]{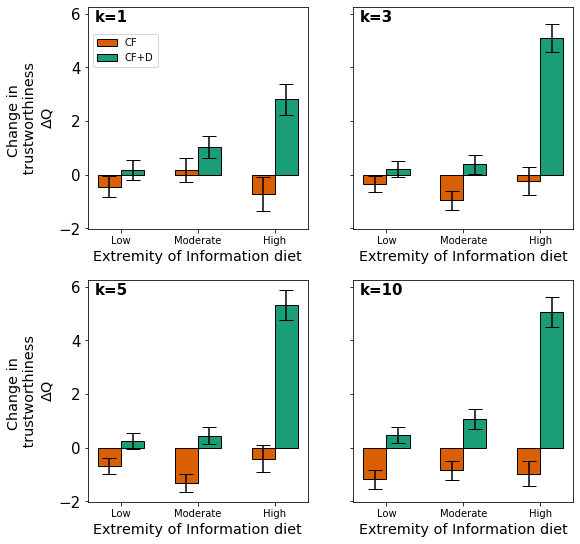}
    \caption{Effect of CF and CF+D versus baseline by
    absolute slant of visited domains (terciles using scores from \citeauthor{BakshyAdamic}) and by length of ranked list $k$. Change in trustworthiness $\Delta Q$ based on scores from NewsGuard~\cite{newsguard}.}
\end{figure}

\begin{figure}
    \includegraphics[width=\textwidth]{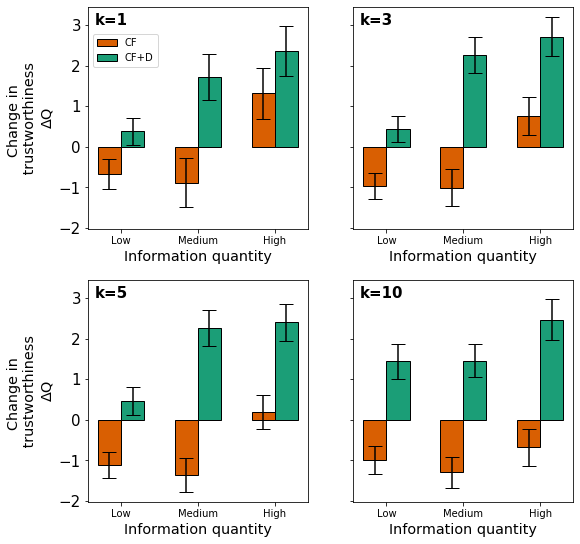}
    \caption{Effect of CF and CF+D versus baseline by total online activity (TF-IDF-transformed pageviews; terciles) and by length of ranked list $k$. Change in trustworthiness $\Delta Q$ based on scores from NewsGuard~\cite{newsguard}.}
\end{figure}

\begin{figure}
    \includegraphics[width=\textwidth]{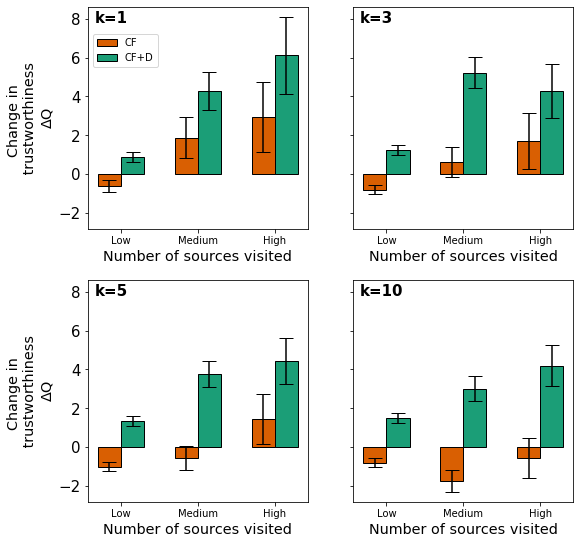}
    \caption{Effect of CF and CF+D versus baseline by distinct number of domains visited (terciles) and by length of ranked list $k$. Change in trustworthiness $\Delta Q$ based on scores from NewsGuard~\cite{newsguard}.}
\end{figure}

\begin{figure}
    \includegraphics[width=\textwidth]{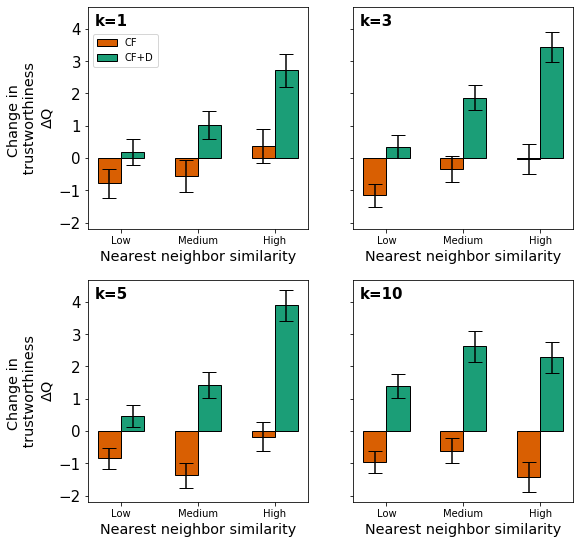}
    \caption{Effect of CF and CF+D versus baseline by average user--user similarity with nearest $n=10$ neighbors in training set (terciles) and by length of ranked list $k$. Change in trustworthiness $\Delta Q$ based on scores from NewsGuard~\cite{newsguard}.}
\end{figure}

\end{document}